\documentclass[prb,twocolumn,showpacs,amsmath,superscriptaddress]{revtex4}
\usepackage[latin1]{inputenc}
\usepackage{color}

\newcommand{\ud}[1]{{#1^{\dagger}}}
\newcommand{\bra}[1]{\left\langle #1\right|}
\newcommand{\ket}[1]{\left| #1\right\rangle}
\newcommand{\braket}[2]{\langle #1|#2\rangle}
\newcommand{\ketd}[1]{|\kern-.066cm| #1\rangle\kern-.1cm\rangle}
\newcommand{\brad}[1]{\langle\kern-.1cm\langle #1|\kern-.066cm|}
\newcommand{\braketd}[2]{\langle #1|\kern-.066cm| #2\rangle\kern-.1cm\rangle}
\newcommand{\braketdd}[2]{\langle\kern-.1cm\langle #1| #2\rangle\kern-.1cm\rangle}

\usepackage{graphicx}
\usepackage{bm}

\begin{document}

\title{Statistics of excitons in quantum dots and the resulting microcavity
emission spectra}
\author{F.~P.~Laussy}
\email{f.p.laussy@sheffield.ac.uk} \affiliation{University of
Sheffield, Department of Physics and Astronomy, Sheffield, S3~7RH,
United Kingdom}
\author{M.~M. Glazov}
\affiliation{A. F. Ioffe Physico-Technical Institute, Russian
Academy of Sciences, 194021 St.~Petersburg, Russia}
\author{A. V. Kavokin}
\affiliation{University of Southampton, Physics and Astronomy
School, Southampton, SO17~1BJ, United Kingdom}
\author{D.~M.~Whittaker}
\affiliation{University of Sheffield, Department of Physics and
Astronomy, Sheffield, S3~7RH, United Kingdom}
\author{G.~Malpuech}
\affiliation{LASMEA, Universit\'e Blaise Pascal, 24, av.~des
Landais, 63~177 Aubi\`ere, France}

\begin{abstract}
  A theoretical investigation is presented of the statistics of
  excitons in quantum dots (QDs) of different sizes. A formalism is
  developed to build the exciton creation operator in a dot from the
  single exciton wavefunction and it is shown how this operator evolves
  from purely fermionic, in case of a small QD, to purely bosonic, in
  case of large QDs.  Nonlinear optical emission spectra of
  semiconductor microcavities containing single QDs are found to
  exhibit a peculiar multiplet structure which reduces to Mollow
  triplet and Rabi doublet in fermionic and bosonic limits,
  respectively.
\end{abstract}

\pacs{71.35.-y, 71.36.+c, 71.10.Pm, 73.21.La}

\maketitle

\section{Introduction}

Semiconductor quantum dots (QDs)~\cite{bimberg98a} are a leading
technology for the investigation of the quantum realm. They offer
exciting possibilities for quantum computation and are important
candidates for the next generation of light emitters. In most
cases, the best control of the states of the confined carriers in
QDs is obtained through coupling to light~\cite{ivchenko05a}. This
light-matter interaction can be considerably enhanced by including
the dot in a microcavity, with pillars~\cite{reithmaier04a},
photonic crystals~\cite{yoshie04a} and microdisks~\cite{peter05a}
being the currently favoured realizations.
References~[\onlinecite{reithmaier04a, yoshie04a, peter05a}]
describe the first reports, in each of these structures, of vacuum
field Rabi splitting, whereby one excitation is transferred back
and forth between the light and the matter fields. This contrasts
with the weak coupling regime previously studied,\cite{gerard98a,
solomon00A} where only quantitative perturbations of the dynamics
occur, such as reductions in the lifetimes of the dot excitations
(Purcell effect).  In the case of strong coupling, however, the
coherent exchange of energy merges the light and matter
excitations into a new entity. This is commonly referred to as an
\emph{exciton-polariton} in semiconductor
physics,~\cite{kavokin03b} with an important example being the
two-dimensional polaritons in planar microcavities, first observed
by Weisbuch \emph{et al.}~\cite{weisbuch92a} In cavity quantum
electrodynamics (cQED), the equivalent concept is the
\emph{dressed state} of atoms by the quantised electromagnetic
field.

In QDs, optical interband excitations create electron-hole pairs
or \emph{excitons}, confined by a three-dimensional potential
which makes their energy spectrum discrete.  If this potential is
much stronger than the bulk exciton binding energy, and if the
size of the dot is smaller than the corresponding exciton Bohr
radius, the Coulomb interaction between electrons and holes can be
considered as a perturbation. For the lowest exciton states, this
is the fermionic limit where the Pauli exclusion principle
dominates. In the opposite limit, if the confining potential is
weak or the size of the dot is much greater than the exciton Bohr
radius, the exciton is quantized as a whole particle. In this
case, the bosonic nature of excitons is expected to prevail over
fermionic nature of individual electrons and holes. An important
question for the description of emission from QDs embedded into
cavities in the strong coupling regime is whether the dot
excitations coupled to light behave like fermions or like bosons.
Here we address the question of which statistics (Bose-Einstein,
Fermi-Dirac or a variation thereof) best describes excitons in
QDs. This is a question which is very topical in view of the
recent experimental achievements, and which has elicited
substantial theoretical works in the past, in connection with the
possibility of exciton Bose condensation.

In this paper we derive the exciton creation operator in a QD
which allows the calculation of nonlinear optical spectra of QDs
in microcavities. The model we develop takes into account the
saturation of the transition due to Pauli exclusion alone and does
not attempt to solve the complex manybody problem which arises
when Coulomb interactions between excitons are included.  Hence
the model is most accurate in describing the departure from ideal
bosonic behaviour in large dots rather than near the fermionic
limit in small dots.  We analyze the dot size effect on the
statistics of excitons and demonstrate the transition from the
fermionic to bosonic regime. To motivate this, we begin by
summarizing how the coupling of light modes with fermionic and
bosonic material excitations differ.

\begin{figure}[htbp]
  \includegraphics[width=\linewidth]{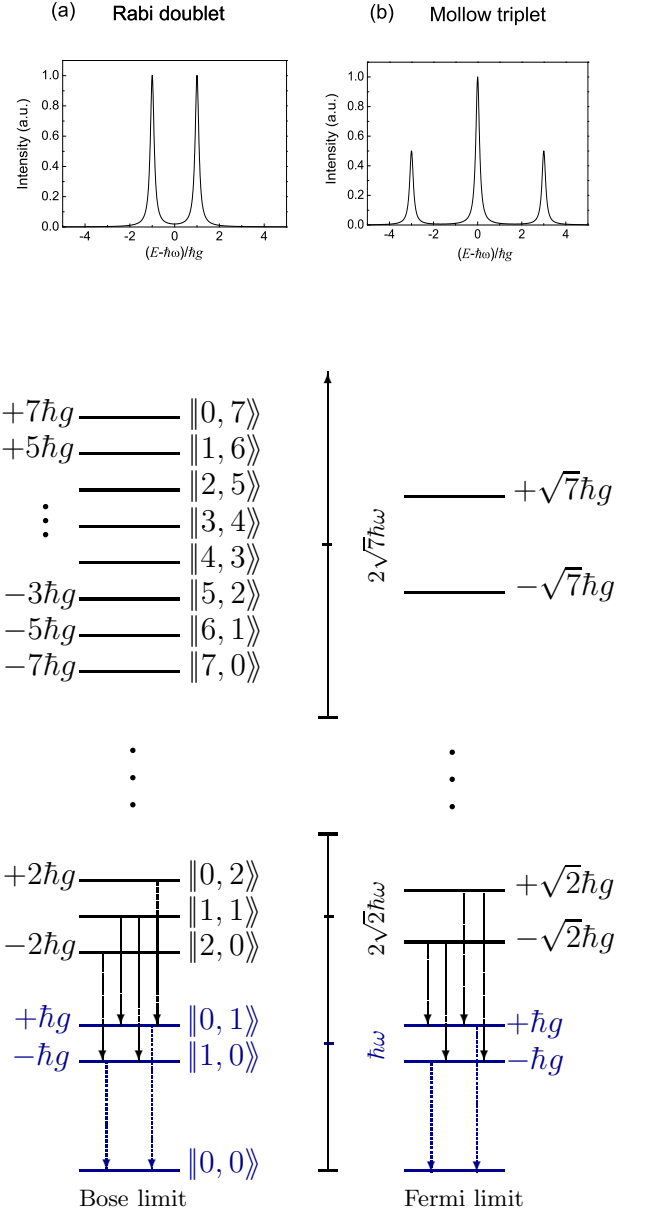}\\
  \vskip.25cm
  \hskip-1.5cm
  \setlength{\unitlength}{1000sp}%
\begingroup\makeatletter\ifx\SetFigFont\undefined%
\gdef\SetFigFont#1#2#3#4#5{%
  \reset@font\fontsize{#1}{#2pt}%
  \fontfamily{#3}\fontseries{#4}\fontshape{#5}%
  \selectfont}%
\fi\endgroup%
\begin{picture}(12765,20727)(-7529,-7951)
\thinlines
{\color[rgb]{0,0,0}\put(361,4214){\line( 1, 0){450}}
}%
{\color[rgb]{0,0,0}\put(361,4180){\line( 1, 0){450}}
}%
{\color[rgb]{0,0,0}\put(361,1325){\line( 1, 0){450}}
}%
{\color[rgb]{0,0,0}\put(361,1291){\line( 1, 0){450}}
}%
\put(-4454,3314){\makebox(0,0)[lb]{\smash{{\SetFigFont{41}{49.2}{\familydefault}{\mddefault}{\updefault}{\color[rgb]{0,0,0}.}%
}}}}
\put(-4454,1964){\makebox(0,0)[lb]{\smash{{\SetFigFont{41}{49.2}{\familydefault}{\mddefault}{\updefault}{\color[rgb]{0,0,0}.}%
}}}}
\put(-4454,2639){\makebox(0,0)[lb]{\smash{{\SetFigFont{41}{49.2}{\familydefault}{\mddefault}{\updefault}{\color[rgb]{0,0,0}.}%
}}}}
\put(-6614,9364){\makebox(0,0)[lb]{\smash{{\SetFigFont{41}{49.2}{\familydefault}{\mddefault}{\updefault}{\color[rgb]{0,0,0}.}%
}}}}
\put(-6614,8669){\makebox(0,0)[lb]{\smash{{\SetFigFont{41}{49.2}{\familydefault}{\mddefault}{\updefault}{\color[rgb]{0,0,0}.}%
}}}}
\put(-6614,9016){\makebox(0,0)[lb]{\smash{{\SetFigFont{41}{49.2}{\familydefault}{\mddefault}{\updefault}{\color[rgb]{0,0,0}.}%
}}}}
\thicklines
{\color[rgb]{0,0,.56}\put(-3104,-7036){\line(-1, 0){2475}}
}%
\thinlines
{\color[rgb]{0,0,.56}\multiput(-4904,-4336)(0.00000,-114.89362){24}{\line( 0,-1){ 57.447}}
\put(-4904,-7036){\vector( 0,-1){0}}
}%
\thicklines
{\color[rgb]{0,0,0}\put(-3104,8039){\line(-1, 0){2475}}
}%
{\color[rgb]{0,0,0}\put(-3104,10739){\line(-1, 0){2475}}
}%
{\color[rgb]{0,0,0}\put(-3104,11639){\line(-1, 0){2475}}
}%
{\color[rgb]{0,0,0}\put(-3104,8939){\line(-1, 0){2475}}
}%
{\color[rgb]{0,0,0}\put(-3104,9839){\line(-1, 0){2475}}
}%
{\color[rgb]{0,0,0}\put(-3104,5339){\line(-1, 0){2475}}
}%
{\color[rgb]{0,0,0}\put(-3104,6239){\line(-1, 0){2475}}
}%
{\color[rgb]{0,0,0}\put(-3104,7139){\line(-1, 0){2475}}
}%
{\color[rgb]{0,0,0}\put(-3104,-1636){\line(-1, 0){2475}}
}%
{\color[rgb]{0,0,0}\put(-3104,-736){\line(-1, 0){2475}}
}%
{\color[rgb]{0,0,0}\put(-3104,164){\line(-1, 0){2475}}
}%
\thinlines
{\color[rgb]{0,0,.56}\multiput(-3779,-3436)(0.00000,-114.28571){32}{\line( 0,-1){ 57.143}}
\put(-3779,-7036){\vector( 0,-1){0}}
}%
\thicklines
{\color[rgb]{0,0,.56}\put(-3104,-3436){\line(-1, 0){2475}}
}%
{\color[rgb]{0,0,.56}\put(-3104,-4336){\line(-1, 0){2475}}
}%
\thinlines
{\color[rgb]{0,0,0}\multiput(-4994,-1636)(0.00000,-56.84211){48}{\line( 0,-1){ 28.421}}
\put(-4994,-4336){\vector( 0,-1){0}}
}%
{\color[rgb]{0,0,0}\multiput(-4094,-736)(0.00000,-57.60000){63}{\line( 0,-1){ 28.800}}
\put(-4094,-4336){\vector( 0,-1){0}}
}%
{\color[rgb]{0,0,0}\multiput(-3644,164)(0.00000,-57.60000){63}{\line( 0,-1){ 28.800}}
\put(-3644,-3436){\vector( 0,-1){0}}
}%
{\color[rgb]{0,0,0}\multiput(-4544,-736)(0.00000,-56.84211){48}{\line( 0,-1){ 28.421}}
\put(-4544,-3436){\vector( 0,-1){0}}
}%
\put(-2924,-4426){\makebox(0,0)[lb]{\smash{{\SetFigFont{12}{14.4}{\familydefault}{\mddefault}{\updefault}{\color[rgb]{0,0,.56}$\ketd{1,0}$}%
}}}}
\put(-2924,-3526){\makebox(0,0)[lb]{\smash{{\SetFigFont{12}{14.4}{\familydefault}{\mddefault}{\updefault}{\color[rgb]{0,0,.56}$\ketd{0,1}$}%
}}}}
\put(-7244,-4426){\makebox(0,0)[lb]{\smash{{\SetFigFont{12}{14.4}{\familydefault}{\mddefault}{\updefault}{\color[rgb]{0,0,.56}$-\hbar g$}%
}}}}
\put(-7199,-3481){\makebox(0,0)[lb]{\smash{{\SetFigFont{12}{14.4}{\familydefault}{\mddefault}{\updefault}{\color[rgb]{0,0,.56}$+\hbar g$}%
}}}}
\put(-2914, 64){\makebox(0,0)[lb]{\smash{{\SetFigFont{12}{14.4}{\familydefault}{\mddefault}{\updefault}{\color[rgb]{0,0,0}$\ketd{0,2}$}%
}}}}
\put(-2914,-1706){\makebox(0,0)[lb]{\smash{{\SetFigFont{12}{14.4}{\familydefault}{\mddefault}{\updefault}{\color[rgb]{0,0,0}$\ketd{2,0}$}%
}}}}
\put(-2924,-826){\makebox(0,0)[lb]{\smash{{\SetFigFont{12}{14.4}{\familydefault}{\mddefault}{\updefault}{\color[rgb]{0,0,0}$\ketd{1,1}$}%
}}}}
\put(-7469,-1681){\makebox(0,0)[lb]{\smash{{\SetFigFont{12}{14.4}{\familydefault}{\mddefault}{\updefault}{\color[rgb]{0,0,0}$-2\hbar g$}%
}}}}
\put(-7469, 74){\makebox(0,0)[lb]{\smash{{\SetFigFont{12}{14.4}{\familydefault}{\mddefault}{\updefault}{\color[rgb]{0,0,0}$+2\hbar g$}%
}}}}
\put(-7469,5249){\makebox(0,0)[lb]{\smash{{\SetFigFont{12}{14.4}{\familydefault}{\mddefault}{\updefault}{\color[rgb]{0,0,0}$-7\hbar g$}%
}}}}
\put(-7469,6149){\makebox(0,0)[lb]{\smash{{\SetFigFont{12}{14.4}{\familydefault}{\mddefault}{\updefault}{\color[rgb]{0,0,0}$-5\hbar g$}%
}}}}
\put(-7469,7049){\makebox(0,0)[lb]{\smash{{\SetFigFont{12}{14.4}{\familydefault}{\mddefault}{\updefault}{\color[rgb]{0,0,0}$-3\hbar g$}%
}}}}
\put(-7514,10649){\makebox(0,0)[lb]{\smash{{\SetFigFont{12}{14.4}{\familydefault}{\mddefault}{\updefault}{\color[rgb]{0,0,0}$+5\hbar g$}%
}}}}
\put(-7514,11549){\makebox(0,0)[lb]{\smash{{\SetFigFont{12}{14.4}{\familydefault}{\mddefault}{\updefault}{\color[rgb]{0,0,0}$+7\hbar g$}%
}}}}
\put(-5579,-7936){\makebox(0,0)[lb]{\smash{{\SetFigFont{9}{10.8}{\familydefault}{\mddefault}{\updefault}{\color[rgb]{0,0,0}Bose limit}%
}}}}
\put(-2924,-7081){\makebox(0,0)[lb]{\smash{{\SetFigFont{12}{14.4}{\familydefault}{\mddefault}{\updefault}{\color[rgb]{0,0,.56}$\ketd{0,0}$}%
}}}}
\put(-2879,5249){\makebox(0,0)[lb]{\smash{{\SetFigFont{12}{14.4}{\familydefault}{\mddefault}{\updefault}{\color[rgb]{0,0,0}$\ketd{7,0}$}%
}}}}
\put(-2884,6144){\makebox(0,0)[lb]{\smash{{\SetFigFont{12}{14.4}{\familydefault}{\mddefault}{\updefault}{\color[rgb]{0,0,0}$\ketd{6,1}$}%
}}}}
\put(-2884,7049){\makebox(0,0)[lb]{\smash{{\SetFigFont{12}{14.4}{\familydefault}{\mddefault}{\updefault}{\color[rgb]{0,0,0}$\ketd{5,2}$}%
}}}}
\put(-2889,7934){\makebox(0,0)[lb]{\smash{{\SetFigFont{12}{14.4}{\familydefault}{\mddefault}{\updefault}{\color[rgb]{0,0,0}$\ketd{4,3}$}%
}}}}
\put(-2884,8824){\makebox(0,0)[lb]{\smash{{\SetFigFont{12}{14.4}{\familydefault}{\mddefault}{\updefault}{\color[rgb]{0,0,0}$\ketd{3,4}$}%
}}}}
\put(-2884,9739){\makebox(0,0)[lb]{\smash{{\SetFigFont{12}{14.4}{\familydefault}{\mddefault}{\updefault}{\color[rgb]{0,0,0}$\ketd{2,5}$}%
}}}}
\put(-2884,10639){\makebox(0,0)[lb]{\smash{{\SetFigFont{12}{14.4}{\familydefault}{\mddefault}{\updefault}{\color[rgb]{0,0,0}$\ketd{1,6}$}%
}}}}
\put(-2889,11534){\makebox(0,0)[lb]{\smash{{\SetFigFont{12}{14.4}{\familydefault}{\mddefault}{\updefault}{\color[rgb]{0,0,0}$\ketd{0,7}$}%
}}}}
\put(3466,3269){\makebox(0,0)[lb]{\smash{{\SetFigFont{41}{49.2}{\familydefault}{\mddefault}{\updefault}{\color[rgb]{0,0,0}.}%
}}}}
\put(3466,1919){\makebox(0,0)[lb]{\smash{{\SetFigFont{41}{49.2}{\familydefault}{\mddefault}{\updefault}{\color[rgb]{0,0,0}.}%
}}}}
\put(3466,2594){\makebox(0,0)[lb]{\smash{{\SetFigFont{41}{49.2}{\familydefault}{\mddefault}{\updefault}{\color[rgb]{0,0,0}.}%
}}}}
\thicklines
{\color[rgb]{0,0,0}\put(361,-7036){\line( 1, 0){450}}
}%
{\color[rgb]{0,0,0}\put(496,-736){\line( 1, 0){225}}
}%
{\color[rgb]{0,0,.56}\put(496,-3886){\line( 1, 0){225}}
}%
\thinlines
{\color[rgb]{0,0,0}\put(586,4214){\vector( 0, 1){8550}}
}%
\thicklines
{\color[rgb]{0,0,0}\put(451,8489){\line( 1, 0){225}}
}%
\thinlines
{\color[rgb]{0,0,0}\put(586,-7036){\line( 0, 1){8325}}
}%
\thicklines
{\color[rgb]{0,0,.56}\put(4951,-7036){\line(-1, 0){2475}}
}%
{\color[rgb]{0,0,.56}\put(4951,-4336){\line(-1, 0){2475}}
}%
{\color[rgb]{0,0,.56}\put(4951,-3436){\line(-1, 0){2475}}
}%
{\color[rgb]{0,0,0}\put(4951,7296){\line(-1, 0){2475}}
}%
{\color[rgb]{0,0,0}\put(4951,9677){\line(-1, 0){2475}}
}%
\thinlines
{\color[rgb]{0,0,.56}\multiput(4276,-3481)(0.00000,-114.28571){32}{\line( 0,-1){ 57.143}}
\put(4276,-7081){\vector( 0,-1){0}}
}%
{\color[rgb]{0,0,0}\multiput(3061,-1371)(0.00000,-56.57534){37}{\line( 0,-1){ 28.288}}
\put(3061,-3436){\vector( 0,-1){0}}
}%
{\color[rgb]{0,0,0}\multiput(3511,-1371)(0.00000,-57.57282){52}{\line( 0,-1){ 28.786}}
\put(3511,-4336){\vector( 0,-1){0}}
}%
{\color[rgb]{0,0,0}\multiput(4456,-101)(0.00000,-56.84564){75}{\line( 0,-1){ 28.423}}
\put(4456,-4336){\vector( 0,-1){0}}
}%
{\color[rgb]{0,0,.56}\multiput(3151,-4381)(0.00000,-114.89362){24}{\line( 0,-1){ 57.447}}
\put(3151,-7081){\vector( 0,-1){0}}
}%
\thicklines
{\color[rgb]{0,0,0}\put(4951,-101){\line(-1, 0){2475}}
}%
\thinlines
{\color[rgb]{0,0,0}\multiput(3961,-101)(0.00000,-57.00855){59}{\line( 0,-1){ 28.504}}
\put(3961,-3436){\vector( 0,-1){0}}
}%
\thicklines
{\color[rgb]{0,0,0}\put(4951,-1371){\line(-1, 0){2475}}
}%
\put(4996,-4516){\makebox(0,0)[lb]{\smash{{\SetFigFont{12}{14.4}{\familydefault}{\mddefault}{\updefault}{\color[rgb]{0,0,.56}$-\hbar g$}%
}}}}
\put(4996,-3616){\makebox(0,0)[lb]{\smash{{\SetFigFont{12}{14.4}{\familydefault}{\mddefault}{\updefault}{\color[rgb]{0,0,.56}$+\hbar g$}%
}}}}
\put(2476,-7936){\makebox(0,0)[lb]{\smash{{\SetFigFont{9}{10.8}{\familydefault}{\mddefault}{\updefault}{\color[rgb]{0,0,0}Fermi limit}%
}}}}
\put(5221,-1501){\makebox(0,0)[lb]{\smash{{\SetFigFont{12}{14.4}{\familydefault}{\mddefault}{\updefault}{\color[rgb]{0,0,0}$-\sqrt{2}\hbar g$}%
}}}}
\put(5221,-151){\makebox(0,0)[lb]{\smash{{\SetFigFont{12}{14.4}{\familydefault}{\mddefault}{\updefault}{\color[rgb]{0,0,0}$+\sqrt{2}\hbar g$}%
}}}}
\put(5221,7229){\makebox(0,0)[lb]{\smash{{\SetFigFont{12}{14.4}{\familydefault}{\mddefault}{\updefault}{\color[rgb]{0,0,0}$-\sqrt{7}\hbar g$}%
}}}}
\put(5176,9614){\makebox(0,0)[lb]{\smash{{\SetFigFont{12}{14.4}{\familydefault}{\mddefault}{\updefault}{\color[rgb]{0,0,0}$+\sqrt{7}\hbar g$}%
}}}}
\put(1981,7544){\rotatebox{90.0}{\makebox(0,0)[lb]{\smash{{\SetFigFont{12}{14.4}{\familydefault}{\mddefault}{\updefault}{\color[rgb]{0,0,0}\small $2\sqrt{7}\hbar\omega$}%
}}}}}
\put(1981,-1726){\rotatebox{90.0}{\makebox(0,0)[lb]{\smash{{\SetFigFont{12}{14.4}{\familydefault}{\mddefault}{\updefault}{\color[rgb]{0,0,0}\small $2\sqrt{2}\hbar\omega$}%
}}}}}
\put(1981,-4291){\rotatebox{90.0}{\makebox(0,0)[lb]{\smash{{\SetFigFont{12}{14.4}{\familydefault}{\mddefault}{\updefault}{\color[rgb]{0,0,.56}\small $\hbar\omega$}%
}}}}}
\end{picture}%
  \caption{(Color online) Energy diagrams of the two limiting cases
    of dressed bosons (left) and fermions (right). In the first case
    the $n$th manifold has constant energy splitting of~$2\hbar g$
    between all states and couples to the $(n-1)$th manifold by
    removal of a quantum of excitation with energy
    $\hbar\omega\pm\hbar g$ which leads to the Rabi doublet, Fig.~(a),
    with splitting~$2\hbar g$.  In the second case, each manifold is
    two-fold with a splitting which increases like a square root.  All
    four transitions are allowed, leading to the Mollow triplet,
    Fig.~(b), for high values of~$n$ when the two middle transitions
    are close in energy.  The distance from the central peak goes
    like~$2\hbar g\sqrt n$ and the ratio of peaks is~$1:2$. The two
    lowest manifolds (in blue) are the same in both cases, making
    vacuum field Rabi splitting insensitive to the
    statistics.}\label{fig:TueAug16164523BST2005}
\end{figure}

The Rabi doublet, with splitting amplitude~$2\hbar g$ as shown on
Fig.~\ref{fig:TueAug16164523BST2005}(a), is well accounted for
theoretically by the coupling with strength~$\hbar g$ of two
quantized oscillators~$a$ and~$b$ both obeying Bose algebra,
\begin{equation}
  \label{eq:WedAug10174624BST2005}
  [a,\ud{a}]=1,
\end{equation}
and equivalently for ~$b$. We shall describe this well known and
elementary case in detail as it provides the foundation for most
of what follows. Neglecting off-resonant terms
like~$\ud{a}\ud{b}$, the hamiltonian reads~\cite{carmichael02a}
\begin{equation}
  \label{eq:TueAug9172340BST2005}
  H=\hbar\omega(\ud{a}a+\ud{b}b)+\hbar g(a\ud{b}+\ud{a}b)\,.
\end{equation}
We assumed degenerate energies~$\hbar\omega$ for the two
oscillators, which will not affect our qualitative results, while
simplifying considerably the analytical expressions.  One
oscillator, say~$a$, describes the light field while the other,
$b$, describes a bosonic matter field. The analysis
of~(\ref{eq:TueAug9172340BST2005}) can be made directly in the
bare state basis~$\ket{i,j}$ with~$i$ excitations in the matter
field and~$j$ in the photonic field, $i$, $j\in\mathbf{N}$. The
value of this approach is that the excitation, loss and dephasing
processes generally pertain to the bare particles. For instance
matter excitations are usually created by an external source
(pumping) and light excitations can be lost by transmission
through the cavity mirror.  This physics is best expressed in the
bare states basis.

On the other hand~(\ref{eq:TueAug9172340BST2005}) assumes a
straightforward expression in the basis of so-called \emph{dressed
  states} which diagonalises the Hamiltonian to read:
\begin{equation}
  \label{eq:TueAug9185727BST2005}
  H=(\hbar\omega-\hbar g)\ud{p}p+(\hbar\omega+\hbar g)\ud{q}q
\end{equation}
where~$\ud{p}$ and~$\ud{q}$ create a coherent superposition of
bare states, respectively in and out of phase:
\begin{equation}
  \label{eq:TueAug9225902BST2005}
  p=(a+b)/\sqrt2,\qquad q=(a-b)/\sqrt2\,.
\end{equation}

For clarity we shall note~$\ketd{i,j}$ the dressed states, i.e.,
the eigenstates of~(\ref{eq:TueAug9185727BST2005}) with~$i$
dressed particles of energy~$\hbar\omega-\hbar g$ and~$j$ of
energy~$\hbar\omega+\hbar g$.  The \emph{manifold} of states
with~$n$ excitations thus reads, in the dressed states basis:
\begin{equation}
  \label{eq:WedAug10003135BST2005}
  \mathcal{H}_n=\{\ketd{i,j}~;~i, j\in\mathbf{N}~\mathrm{with}~i+j=n\}
\end{equation}
Its energy diagram appears on the left of
Fig.~\ref{fig:TueAug16164523BST2005} for manifolds with zero
(vacuum), one, two and seven excitations. When an excitation
escapes the system while in manifold~$\mathcal{H}_n$, a transition
is made to the neighbouring manifold~$\mathcal{H}_{n-1}$ and the
energy difference is carried away, either by the leaking out of a
cavity photon, or through exciton emission into a radiative mode
other than that of the cavity, or a non-radiative process.  The
detailed analysis of such processes requires a dynamical study,
but as the cavity mode radiation spectra can be computed with the
knowledge of only the energy level diagrams, we shall keep our
analysis to this level for the present work. The important feature
of this dissipation is that, though such processes involve~$a$
or~$b$ (rather than~$p$ or~$q$), they nevertheless still result in
removing one excitation out of one of the oscillators. Hence only
transitions from~$\ketd{i,j}$ to~$\ketd{i-1,j}$ or~$\ketd{i,j-1}$
are allowed, bringing away, respectively,~$\hbar\omega+\hbar g$
and~$\hbar\omega-\hbar g$ of energy, accounting for the so-called
Rabi doublet (provided the initial $i$ and~$j$ are nonzero in
which case only one transition is allowed). From the algebraic
point of view, this of course follows straightforwardly
from~(\ref{eq:TueAug9225902BST2005}) and orthogonality of the
basis states. Physically it comes from the fact that, as in the
classical case, the coupled system acts as two independent
oscillators vibrating with frequencies $\omega\pm g$. In the case
of vacuum field Rabi splitting, a single excitation is shared
between the two fields, and so the manifold~$\mathcal{H}_1$ is
connected to the single line of the vacuum manifold. In this case
there is obviously no possibility beyond a doublet.

Different physics occurs when the excitations of the material are
described by fermionic rather than bosonic statistics. In the case
of cavity QED the material is usually a beam of atoms passing
through the cavity, and a single excitation is the independent
responses of the atoms to the light field excitation. The simplest
situation is that of a dilute atomic beam where a single atom
(driven at resonance so that it appears as a two-levels system) is
coupled to a Fock state of light with a large number of photons.
This case is described theoretically by the Jaynes-Cummings
model~\cite{jaynes63a}, in which $a$ (the radiation field) remains
a Bose operator but~$b$ becomes a fermionic operator which
describes two-levels systems, $\ud{b}\rightarrow \sigma_+$ with~:
\begin{equation}
  \label{eq:WedAug10025855BST2005}
  \sigma_+=
  \begin{pmatrix}
    0&1\\
    0&0
  \end{pmatrix}.
\end{equation}

Then the atom must be in either the ground or the excited state,
allowing for the manifolds
\begin{equation}
  \label{eq:WedAug10025024BST2005}
  \mathcal{H}_n=\{\ket{0,n}, \ket{1, n-1}\}
\end{equation}
provided that~$n\ge 1$. The associated energy diagrams appear on
the right of Fig.~\ref{fig:TueAug16164523BST2005}, with two states
in each manifold (in our conventions~$\ket{0,n}$ refers to the
bare states with atom in ground state and~$n$ photons,
while~$\ket{1, n-1}$ has atom in excited state and~$n-1$ photons).
For the resonant condition, where the two-levels transition
matches the cavity photon energy, the dressed states for this
manifold are split by an energy $\sqrt{n} \hbar g$.  In the
general case, all four transitions between the states in manifolds
$\mathcal{H}_n$ and $\mathcal{H}_{n-1}$ are possible, which
results in a quadruplet. It is hard to resolve this quadruplet,
but it has been done in Fourier transform of time resolved
experiments~\cite{brune96a}. It is simpler to consider
photoluminescence directly under continuous excitation at high
intensity (where the fluctuations of particles number have little
effect). In this case, with~$n\gg 1$, the two intermediate
energies are almost degenerate and a triplet is obtained with its
central peak being about twice as high as the two satellites.
This is the Mollow triplet of resonance
fluorescence~\cite{mollow69a}.

Thus we are faced with two limiting cases, one is a pure
\emph{bosonic} limit with equally spaced dressed states resulting
in the linear Rabi doublet, the other the pure \emph{fermionic}
limit with pairs of dressed states of increasing splitting within
a manifold but decreasing energy difference between two successive
manifolds, giving rise to the Mollow triplet.  In many of the
strong coupling experiments conducted so far, and in all the
reports concerning semiconductor QDs, only one single excitation
is exchanged coherently, so that the states are dressed by the
vacuum of the electromagnetic field, resulting in the Rabi
doublet. However, at this level of excitation, there is complete
agreement between the bosonic and fermionic models, with both
providing a good description of the experimental observations. The
prospect of stronger pumping, with more than one excitation shared
between the two fields, makes it important to understand whether a
realistic semiconductor QD will correspond to the bosonic or
fermionic case, or something intermediate between the two.

We review and discuss some of the more significant achievements in
this field in section~\ref{sec:excitons}.  In
section~\ref{sec:formalism} we lay down a general formalism for
building the exciton creation operator. In
section~\ref{sec:limits} we study two limiting cases which
resemble Bose and Fermi statistics. We show how in the general
case the luminescence behaviour interpolates between these two
limits which we have already discussed, and calculate the second
order coherence of the emitted light. In
section~\ref{sec:multiplets} we draw the experimental consequences
of the various statistics and discuss how the spectra obtained
allow a qualitative understanding of how excitations distribute
themselves in the excitonic field.  In the final section we
conclude and discuss briefly ongoing work to refine the modelling
of the excitonic wavefunction.

\section{Excitons as quasi-particles}\label{sec:excitons}

The generic optical excitation in an intrinsic semiconductor is
the electron-hole pair. In bulk, the two oppositely charged
particles can be strongly correlated by the Coulomb interaction
and bound as hydrogenic states (Wannier-Mott exciton). Although
finding binding energies and wavefunctions for the single exciton
case is a difficult problem in various
geometries,\cite{dandrea90a, que92a, he05a} as far as the vacuum
coupling limit is concerned, the exciton field operator which
links the two manifolds~$\mathcal{H}_1$ and vacuum always assumes
the simple form~$\sigma_+$ regardless of the details of the
exciton.  As a particle constituted of two fermions, the exciton
is commonly regarded as a boson, from consideration of the angular
momentum addition rules and the spin-statistics theorem.  For a
single particle, this is an exact statement, albeit a trivial one.
It is however generally agreed to hold at small
densities~\cite{keldysh68a}.  In this case the Rabi doublet is
obtained, as observed experimentally~\cite{reithmaier04a,
yoshie04a, peter05a}.

At higher excitation power, the problem assumes considerable
complexity as well as fundamental importance for physical
applications. Already at the next higher excitation---with one
more electron, hole or electron-hole pair added to the first
exciton---the situation offers rich and various phenomena both in
weak~\cite{besombes02a, patton03a} and strong~\cite{ivanov98a}
coupling, owing to the underlying complexity of the excitonic
states. In this work we shall be concerned with resonant optical
pumping, so that excitations are created in pairs and the system
always remains electrically neutral. We shall describe as an
\emph{exciton} any state of an electron-hole pair, whether it is
an atom-like 1$s$ state or has both particles independently
quantised,~\cite{que92a} and in a more general sense we shall also
use the term for any combination of particles which takes part in
the excitonic phase.  Indeed the excitonic phase with more than
one pair requires at the most accurate level a description in
terms of an excitonic complex, e.g., in term of
biexcitons/bipolaritons for strong coupling of two electrons and
two holes with light~\cite{ivanov98a}. This comes from the Coulomb
interaction which links all charge carriers together and, in a
most fundamental way, also from the antisymmetry of the
wavefunction which demands a sign change whenever two identical
fermions (electrons or holes) are swapped in the system. However
in some configurations, especially in planar cavities, a widely
accepted hypothesis of bosonic behavior of excitons and the
derived polaritons has been investigated for effects such as the
exciton boser~\cite{imamoglu96a}, polariton
amplifier~\cite{savvidis00a, ciuti00a, rubo03a} and polariton
lasers~\cite{deng02a, laussy04c}.  The internal structure of the
exciton which gives rise to both deviations from the
Bose-statistics and interactions of the electron-hole pairs is
then expressed as an effective repulsive force in a bosonised
Hamiltonian (due to Coulomb interaction and Pauli effect in the
form of phase-space filling or exchange
interaction~\cite{imamoglu98a, ciuti98a}).

This bosonic approach for excitons met early opposition in favour
of an analysis in the electron-hole
basis~\cite{kira97a,bentabou01a,bentabou03a}. Combescot and
co-workers investigated the possibility of bosonisation of
excitons~\cite{combescot01a, combescot02a, combescot02b,
combescot04b} and concluded against it. They point out its
internal inconsistency, as the same interaction binds the
underlying fermions, and therefore defines the exciton, while also
being responsible for exciton-exciton scattering; this is
inconsistent with the indistinguishability of the particles. These
authors introduce the ``proteon'' as the paradigm for Bose-like
composite particle and propose a formalism (``commutation
technics'') which essentially relies on evaluating quantum
correlators in the fermion basis with operators linked through the
single exciton wave-function.  The importance of Fermi statistics
of the underlying constituents has also been pointed out by
Rombouts and co-workers in connection to atomic and excitonic
condensates~\cite{rombouts02a, rombouts05a}.  In both cases the
composite ``boson'' creation operator reads
\begin{equation}
  \label{eq:WedAug10171604BST2005}
\ud{B}=\sum_\mathbf{k}\phi_\mathbf{k}\ud{\sigma_{\mathbf{k}}}\ud{\varsigma_{-\mathbf{k}}}
\end{equation}
in term of~$\sigma_\mathbf{k}$, $\varsigma_\mathbf{k}$ the fermion
operators
\begin{equation}
  \label{eq:WedAug10174033BST2005}
  \{\sigma_{\mathbf{k}},\ud{\sigma_{\mathbf{k}'}}\}
=\delta_{\mathbf{k},\mathbf{k}'}, \qquad
\{\sigma_{\mathbf{k}},\sigma_{\mathbf{k}'}\}=0
\end{equation}
(same for~$\varsigma_{\mathbf{k}}$), respectively for an electron
and hole (or proton in the atomic case~\cite{rombouts05a}) of
momentum~$\mathbf{k}$. This operator creates excitons with the
center-of-mass momentum~$\mathbf{K}=\mathbf{0}$ in a system with
translational invariance. It was appreciated long
ago\cite{keldysh68a} that the operators $\ud{B}$ and $B$ are
no-longer exact bosonic operators, but instead satisfy,
\begin{equation}\label{eq:commutator:approx}
  [B, \ud{B}] = 1 - D,
\end{equation}
where $D$ is a nonzero operator, though with small matrix elements
at low exciton densities.

Our approach is based on a definition similar to
Eq.~(\ref{eq:WedAug10171604BST2005}) for the exciton operator.
Instead of analysing the deviations in the commutation
relationship (\ref{eq:commutator:approx}) we derive the matrix
elements of the operator $\ud{B}$. The direct analysis of these
matrix elements allows us to trace departures from Bose-statistics
and to investigate the transitions between bosonic and fermionic
behaviours of excitons in QDs. We shall compare our results with
those of Refs.~[\onlinecite{combescot01a, combescot02a,
combescot02b, combescot04b,rombouts02a, rombouts05a}].  For
convenience, we will refer to these two sets of publications
through the names of their first authors, keeping in mind that
they are co-authored papers, as listed in the references.

\section{Formalism}\label{sec:formalism}

We consider a QD, which localises the excitation in real space.
Thus our main departure from Combescot and Rombouts is that our
exciton creation operator, $\ud{X}$, is expressed in real space
and without the zero center-of-mass momentum restriction,
\begin{equation}
  \label{eq:WedAug10193047BST2005}
  \ud{X}=\sum_{n_e,n_h}C_{n_e,n_h}\ud{\sigma_{n_e}}\ud{\varsigma_{n_h}}
\end{equation}
where $\sigma_{n_e}$ and~$\varsigma_{n_h}$ are fermion creation
operators, cf.~(\ref{eq:WedAug10174033BST2005}), for an electron
and a hole in state~$\ket{\varphi^e_{n_e}}$
and~$\ket{\varphi^h_{n_h}}$, respectively:
\begin{equation}
  \label{eq:WedAug10193429BST2005}
\ud{\sigma_{n_e}}\ket{0}=\ket{\varphi^e_{n_e}},\quad\ud{\varsigma_{n_h}}\ket{0}=\ket{\varphi^h_{n_h}},
\end{equation}
with~$\ket{0}$ denoting both the electron and hole vacuum fields.
We carry out the analysis in real space with the set of basis
wavefunctions
\begin{equation}
  \label{eq:WedAug10193744BST2005}
\varphi^e_{n_e}(\mathbf{r}_e)=\braket{\mathbf{r}_e}{\varphi^e_{n_e}}\quad\mathrm{and}\quad\varphi^h_{n_h}(\mathbf{r}_h)=\braket{\mathbf{r}_h}{\varphi^h_{n_h}}
\end{equation}
with~$\mathbf{r}_e$ and~$\mathbf{r}_h$ the positions of the
electron and hole, respectively. Subscripts $n_e$ and $n_h$ are
multi-indices enumerating all quantum numbers of electrons and
holes. The specifics of the three-dimensional confinement
manifests itself in the discrete character of $n_e$ and $n_h$
components.

We restrict our considerations to the direct band semiconductor
with non-degenerate valence band. Such a situation can be
experimentally achieved in QDs formed in conventional III-V or
II-VI semiconductors, where the light-hole levels lie far, in
energy, heavy-hole ones due to the effects of strain and size
quantization along the growth axis\cite{ivchenko05a}. Therefore,
only electron-heavy hole excitons need to be considered. Moreover,
we will neglect the spin degree of freedom of the electron-hole
pair and assume all carriers to be spin polarized. This can be
realized by pumping the system with light of definite circular
polarization, as spin-lattice relaxation is known to be very
inefficient in QDs.

The (single) exciton wavefunction~$\ket{\varphi}$ results from the
application of~$\ud{X}$ on the vacuum. In real space coordinates:
\begin{equation}
  \label{eq:WedAug10225813BST2005}
\braket{\mathbf{r}_e,\mathbf{r}_h}{\varphi}=\varphi(\mathbf{r}_e,\mathbf{r}_h)=\sum_{n_e,n_h}C_{n_e,n_h}\varphi^e_{n_e}(\mathbf{r}_e)\varphi^h_{n_h}(\mathbf{r}_h).
\end{equation}

At this stage we do not specify the wavefunction (that is, the
coefficients~$C_{n_e,n_h}$), which depends on various factors such
as the dot geometry, electron and hole effective masses and
dielectric constant. Rather, we consider the~$n$ excitons state
which results from successive excitation of the system
through~$\ud{X}$:
\begin{equation}
  \label{eq:WedAug10230402BST2005}
  \ket{\Psi_n}=(\ud{X})^n\ket{0},
\end{equation}
We later discuss in more details what approximations are being
made here. For now we proceed by normalising this wavefunction
\begin{equation}
  \label{eq:WedAug10230706BST2005}
  \ket{n}={1\over\mathcal{N}_n}\ket{\Psi_n}
\end{equation}
where, by definition of the normalization constant
\begin{equation}
  \label{eq:WedAug10230757BST2005}
  \mathcal{N}_n=\sqrt{\braket{\Psi_n}{\Psi_n}}.
\end{equation}

The creation operator~$\ud{X}$ can now be obtained explicitly.  We
call~$\alpha_n$ the non-zero matrix element which lies below the
diagonal in the excitons representation:
\begin{equation}
  \label{eq:WedAug10231252BST2005}
  \alpha_n = \langle n |\ud{X}|n-1\rangle\,,
\end{equation}
which, by comparing
Eqs.~(\ref{eq:WedAug10230402BST2005}--\ref{eq:WedAug10231252BST2005})
turns out to be
\begin{equation}
  \label{eq:WedAug10231324BST2005}
  \alpha_n = \frac{\mathcal N_n}{\mathcal N_{n-1}}\,.
\end{equation}

We now undertake to link~$\alpha_n$ with the
coefficients~$C_{n_e,n_h}$, which assume a specific form only when
the system itself has been characterised.  The general
relationship is more easily obtained in the real space than with
the operator representation~(\ref{eq:WedAug10193047BST2005}).
Indeed the non-normalized $n$-excitons wavefunction assumes the
simple form of a Slater determinant
\begin{equation}
  \label{eq:WedAug10232444BST2005}
\Psi_n(\mathbf{r}_{e_1},\ldots,\mathbf{r}_{e_n},\mathbf{r}_{h_1},\ldots,\mathbf{r}_{h_n})=\det_{1\leq
i,j \leq n}{[\varphi(\mathbf{r}_{e_i}, \mathbf{r}_{h_j})]},
\end{equation}
explicitly ensuring the antisymmetry of~$\Psi_n$ upon exchange of
two identical fermions (holes and electrons), as results from the
anticommutation rule~(\ref{eq:WedAug10174033BST2005}) in the $n$th
power of operator~$\ud{X}$, cf.~(\ref{eq:WedAug10193047BST2005}).

The determinant can be computed explicitly, by expansion of its
minors which results in the recurrent relation
\begin{equation}
  \label{eq:ThuAug11001515BST2005}
  \mathcal N_n^2 = {1\over n}\sum_{m=1}^n (-1)^{m+1} \beta_m \mathcal N_{n-m}^2
\prod_{j=0}^{m-1} (n-j)^2,
\end{equation}
with~$\mathcal{N}_0=1$ and $\beta_m$ the irreducible $m$-excitons
overlap integrals, $1\le m\le n$:
\begin{widetext}
\begin{equation}
  \label{eq:ThuAug11001646BST2005}
\beta_m=\int\left(\prod_{i=1}^{m-1}\varphi^*(\mathbf{r}_{e_i},\mathbf{r}_{h_i})\varphi(\mathbf{r}_{e_i},\mathbf{r}_{h_{i+1}})\right)\varphi^*(\mathbf{r}_{e_m},\mathbf{r}_{h_m})\varphi(\mathbf{r}_{e_m},\mathbf{r}_{h_1})\,d\mathbf{r}_{e_1}\dots
d\mathbf{r}_{e_m}d\mathbf{r}_{h_1}\dots d\mathbf{r}_{h_m}
\end{equation}
\end{widetext}
The determinant can also be solved by direct combinatorial
evaluation, counting all combinations which can be factored out as
products of~$\beta_m$. The result expressed in this way reads:
\begin{equation}
  \label{eq:ThuAug11001945BST2005}
\mathcal{N}_n^2=\sum_{\eta=1}^{p(n)}C_n(\eta)\prod_{m=1}^N\beta_m^{\nu_{\eta}(m)}
\end{equation}
where~$p(n)$ is the partition function of~$n$ (number of ways to
write~$n$ as a sum of positive integers, i.e., as an integer
partition of~$n$) and~$\nu_\eta(i)$ is the number of times
that~$i\in\mathbf{N}$ appears in the~$\eta^\mathrm{th}$ partition
of~$n$. The coefficients~$C_n$ read:
\begin{equation*}
    C_n(\eta)\equiv
n!(-1)^{n+\sum_{m=1}^n\nu_\eta(m)}\prod_{\{n_i\}}{n-\sum_{j=1}^{i-1}n_j\choose
n_i}{(n_i-1)!\over\nu_\eta(i)!}
\end{equation*}
where the product is taken over the integers $n_i$ which enter in
the $\eta^\mathrm{th}$ partition of~$n$, i.e., $n=\sum_i n_i$.

The procedure to calculate the matrix elements of the creation
operator is as follows: One starts from the envelope function
$\varphi(\mathbf{r}_e, \mathbf{r}_h)$ for a single exciton.  Then
one calculates all overlap integrals $\beta_m$ as given
by~(\ref{eq:ThuAug11001646BST2005}), for~$1\le m\le n$ where~$n$
is the highest manifold to be accessed.  Then the norms can be
computed, the more practical way being recursively
with~(\ref{eq:ThuAug11001515BST2005}).  Finally the matrix
elements $\alpha_n$ are obtained as the successive norms ratio,
cf.~(\ref{eq:WedAug10231252BST2005}). Once $\alpha_n$ are known
the emission spectra can be calculated with ease. We note here
that the numerical computation of the $\beta_m$ and $\alpha_n$
values needs to be carried out with great care. The cancellation
of the large numbers of terms involved in
Eq.~(\ref{eq:ThuAug11001515BST2005}) requires a very
high-precision evaluation of $\beta_m$.

Although derived from the formulation in real
space~(\ref{eq:WedAug10232444BST2005}), the recurrent
relation~(\ref{eq:ThuAug11001515BST2005}), or its analytical
solution~(\ref{eq:ThuAug11001945BST2005}), is a property of
fermion pairs, so it applies to the fermionic operators
in~(\ref{eq:WedAug10171604BST2005}) as well. The core of the
mathematical results contained in these two expressions has in
fact been obtained by Combescot~\cite{combescot01a} through direct
evaluation with the operator algebra involved in~$\ud{B}$. The
only difference with her approach and ours is that her
corresponding quantity~$\beta_m$ (which she notes~$\sigma_m$)
appears as a series in the (reciprocal space) wavefunction
\begin{equation}
  \label{eq:ThuAug11003437BST2005}
  \sigma_m=\sum_{\mathbf{k}}|\phi_\mathbf{k}|^{2m}
\end{equation}
as opposed to the overlap
integral~(\ref{eq:ThuAug11001646BST2005}).

\begin{figure}[htbp]
  \includegraphics[width=\linewidth]{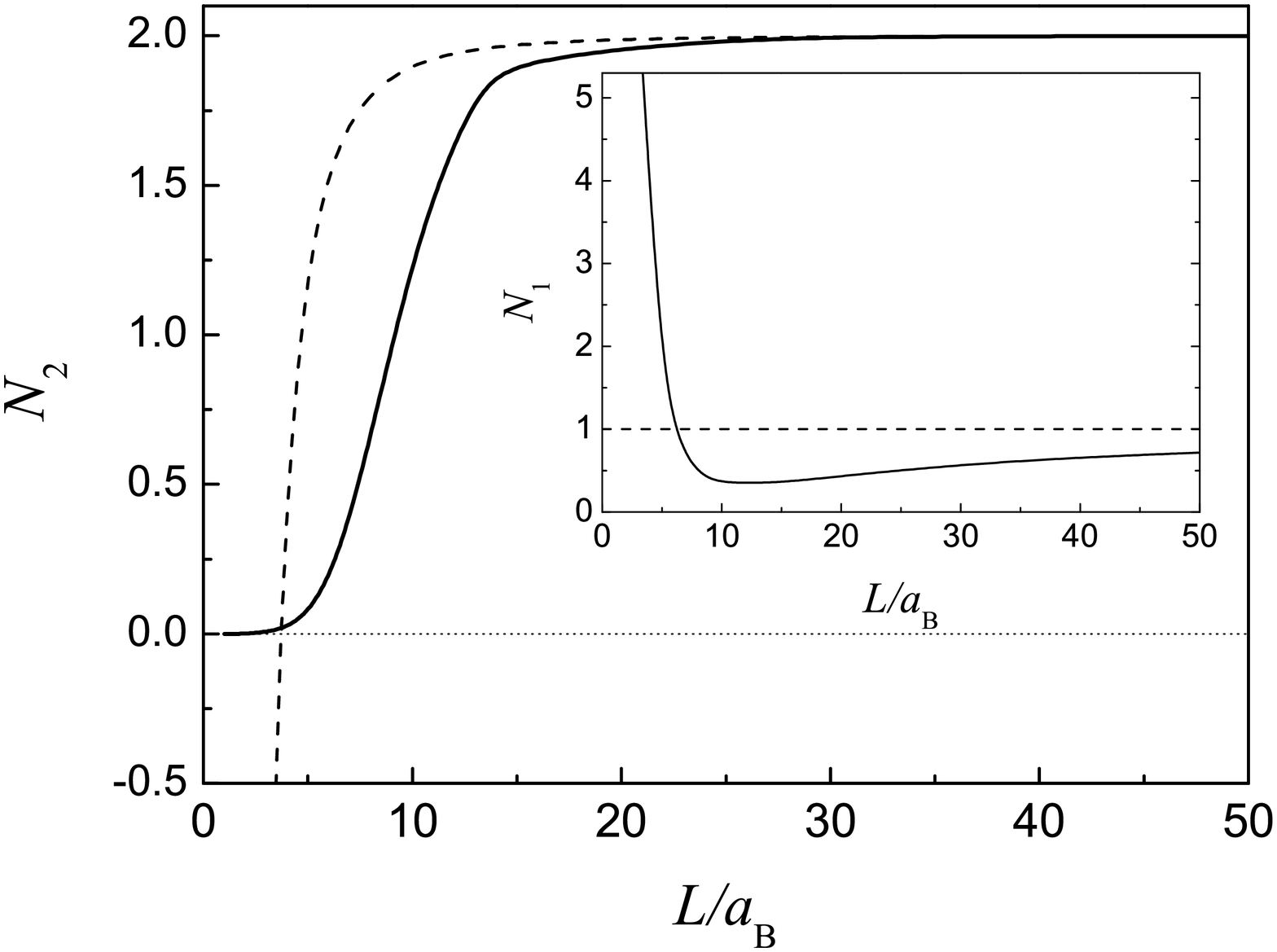}\\
  \caption{Normalization constant $\mathcal N_2$ vs.~ratio of crystal
    size $L$ to 3D exciton Bohr radius $a_{\mathrm B}$.  Dashed line
    shows the result using the approximated
    expression~(\ref{eq:ThuAug11011718BST2005}) for $\sigma_m$ and
    solid line using the exact evaluation of the sum on the mesh in
    reciprocal space. The negative value of the norm used as a
    criterion for boson wavefunction breakdown in
    Ref.~[\onlinecite{combescot01a}] is an artifact of this
    approximation. With the genuine wavefunction, our procedure yields
    a non-arbitrary way to consider how the wavefunction vanishing
    norm affects the bosonic character of the excitation.  Inset
    displays $\mathcal N_1$ vs.~$L/a_{\mathrm B}$ computed on mesh
    (solid) and according to Eq.~(\ref{eq:ThuAug11011718BST2005})
    (dashes).}\label{fig:ThuSep29125153BST2005}
\end{figure}

With such a simple expression as~(\ref{eq:ThuAug11003437BST2005}),
Combescot \emph{et al.} have been able to obtain approximate
analytical forms for 1$s$ states of excitons in both~3D and~2D.
The sum over reciprocal space is approximated as an integral in
the continuum limit, and since $\phi_\mathbf{k}$ depends
on~$k=|\mathbf{k}|$ only, this becomes
$\sum_\mathbf{k}|\phi_\mathbf{k}|^{2m}\rightarrow\int
\mathcal{V}_k|\phi_k|^{2m}dk$ with~$\mathcal{V}_k=4\pi k^2$
or~$2\pi k$, respectively, yielding~\cite{combescot01a}
\begin{align}
  \label{eq:ThuAug11011718BST2005}
  \sigma_m&=z^{m-1}16{(8m-5)!!\over(8m-2)!!}&\mathrm{in~3D},\\
  \sigma_m&={2y^{m-1}\over3m-1}&\mathrm{in~2D},
\end{align}
with~$z$ and~$y$ dimensionless parameters involving the ratio of
Bohr radius~$a_\mathrm{B}$ to the size of the system. In
Ref.~[\onlinecite{combescot01a}], this continuous approximation is
used even when the system size becomes so small that the
quantization becomes noticeable, as a result of which a negative
norm is obtained even for a well defined wavefunction.  However,
the authors interpret incorrectly this negative norm as the result
of an unphysical exciton wavefunction and thus the breakdown of
the bosonic picture. To demonstrate this, we have computed
numerically the values of $\sigma_m$ by direct summation of
Eq.~(\ref{eq:ThuAug11003437BST2005}) and used them to evaluate
normalization constants $\mathcal N_n$. Some results of the
computation are shown in Fig.~\ref{fig:ThuSep29125153BST2005}. The
inset shows the dependence of $\mathcal N_1 \equiv \sigma_1$
vs.~the ratio of crystal size $L$ to 3D exciton Bohr radius
$a_{\mathrm B}$. The continuous approximation
Eq.~(\ref{eq:ThuAug11011718BST2005}) gives $\sigma_1=1$, shown by
the dashed line in the inset. The solid line displays the value
of~$\sigma_1$ computed exactly
with~(\ref{eq:ThuAug11003437BST2005}): it approaches~$1$ when
$L/a_{\mathrm B} \to \infty$ but departs strongly from~1 as the
size of the system reduces, since the real wavefunction changes as
the system shrinks about it. At the very least this needs to be
taken into account by correcting the normalization constant
of~$\phi_\mathbf{k}$, as otherwise unphysical result may arise.
The main part of the figure shows the two-excitons normalization
$\mathcal N_2$ calculated by both methods. We find that for all
$L/a_{\rm B}$, the exact $\mathcal N_2$ never becomes negative in
clear constrast with the behavior obtained with the approximated
value~(\ref{eq:ThuAug11011718BST2005}). Hence it is not possible
to use the appearance of a  negative norm as a criterion for
bosonic breakdown.

\section{Limiting cases}\label{sec:limits}

We shall not attempt in this paper to go through the lenghty and
complicated task of the numerical calculation of the exciton
creation operator matrix elements for a realistic QD. Rather, we
consider here a model wavefunction which can be integrated
analytically and illustrates some expected typical behaviours.
Before this, we comment briefly on the limiting cases of
Bose-Einstein and Fermi-Dirac statistics and how they can be
recovered in the general setting of this section.

This will be made most clear through consideration of the explicit
case of two excitons ($n=2$). Then the wavefunction reads
\begin{multline}
  \label{eq:WedAug10233214BST2005}
\Psi_2(\mathbf{r}_{e_1},\mathbf{r}_{e_2},\mathbf{r}_{h_1},\mathbf{r}_{h_2})=\varphi(\mathbf{r}_{e_1},\mathbf{r}_{h_1})\varphi(\mathbf{r}_{e_2},\mathbf{r}_{h_2})\\
-\varphi(\mathbf{r}_{e_1},\mathbf{r}_{h_2})\varphi(\mathbf{r}_{e_2},\mathbf{r}_{h_1})
\end{multline}
with its normalization constant~(\ref{eq:WedAug10230757BST2005})
readily obtained as
\begin{equation}
  \label{eq:ThuAug11022947BST2005}
  \mathcal N_2^2 = \int |\Psi_2(\mathbf{r}_{e_1}, \mathbf{r}_{e_2},
\mathbf{r}_{h_1}, \mathbf{r}_{h_2})|^2 d\mathbf{r}_{e_1} \ldots
d\mathbf{r}_{h_2}
  =2-2\beta_2
\end{equation}
where $\beta_2$, the two-excitons overlap integral, reads
explicitly
\begin{multline}
  \label{eq:ThuAug11023421BST2005}
  \beta_2 =\int \varphi(\mathbf{r}_{e_1}, \mathbf{r}_{h_1})
\varphi(\mathbf{r}_{e_2}, \mathbf{r}_{h_2})\\
  \varphi(\mathbf{r}_{e_1}, \mathbf{r}_{h_2})\varphi(\mathbf{r}_{e_2},
\mathbf{r}_{h_1}) d\mathbf{r}_{e_1} \ldots d\mathbf{r}_{h_2}.
\end{multline}

This integral is the signature of the composite nature of the
exciton. The minus sign in~(\ref{eq:ThuAug11022947BST2005})
results from the Pauli principle: two fermions (electrons and
holes) cannot occupy the same state.  Assuming
$\varphi(\mathbf{r}_e,\mathbf{r}_h)$ is normalised, $\mathcal
N_1=1$, so according to~(\ref{eq:WedAug10231324BST2005}),
\begin{equation}
  \label{eq:ThuAug11023909BST2005}
  \alpha_2 = \sqrt{2 - 2\beta_2}.
\end{equation}
Since $0\leq\beta_2\leq 1$ this is smaller than or equal to
$\sqrt{2}$, the corresponding matrix element of a true boson
creation operator. This result has a transparent physical meaning:
since two identical fermions from two excitons cannot be in the
same quantum state, it is ``harder'' to create two real excitons,
where underlying structure is probed, than two ideal bosons. We
note that if $L$ is the QD lateral dimension, $\beta_2\sim
(a_{\mathrm B}/L)^2 \ll 1$ when $L\gg a_{\mathrm B}$. Thus in
large QDs the overlap of excitonic wavefunctions is small, so
$\alpha_2\approx \sqrt{2}$ and the bosonic limit is recovered. On
the other hand, in a small QD, where Coulomb interaction is
unimportant compared to the dot potential confining the carriers,
the electron and hole can be regarded as quantized separately:
\begin{equation}
  \label{eq:ThuAug11025623BST2005}
\varphi(\mathbf{r}_e,\mathbf{r}_h)=\varphi^e(\mathbf{r}_e)\varphi^h(\mathbf{r}_h)
\end{equation}
In this case all~$\beta_m=1$ and subsequently all~$\alpha_m=0$ at
the exception of~$\alpha_1=1$. This is the fermionic limit
where~$\ud{X}$ maps to the Pauli matrix~$\sigma_+$,
cf.~(\ref{eq:WedAug10025855BST2005}).

We now turn to the general case of arbitrary sized QDs,
interpolating between the (small) fermionic and (large) bosonic
limits.  We assume a Gaussian form for the wavefunction which
allows to evaluate analytically all the required quantities. As
numerical accuracy is not the chief goal of this work we further
assume in-plane coordinates~$x$ and~$y$ to be uncorrelated to ease
the computations. The wavefunction reads:
\begin{equation}
  \label{eq:MonNov14233031GMT2005}
\varphi(\mathbf{r}_e,\mathbf{r}_h)=\mathcal{C}\exp(-\gamma_e\mathbf{r}_e^2-\gamma_h\mathbf{r}_h^2-\gamma_{eh}\mathbf{r}_e\cdot\mathbf{r}_h)
\end{equation}
properly normalized with
\begin{equation}
  \label{eq:ThuAug11141718BST2005}
  \mathcal{C}={\sqrt{4\gamma_e\gamma_h-\gamma_{eh}^2}\over\pi}
\end{equation}
provided that~$\gamma_{eh}\in[-2\sqrt{\gamma_e\gamma_h},0]$
with~$\gamma_e$, $\gamma_h\ge0$. The $\gamma$ parameters allow to
interpolate between the large and small dot limits within the same
wavefunction.  To connect these parameters~$\gamma_e$, $\gamma_h$
and $\gamma_{eh}$ to physical quantities,
(\ref{eq:MonNov14233031GMT2005}) is regarded as a trial
wavefunction which is to minimize the
hamiltonian~$H_{\mathrm{QD}}$ confining the electron and hole in a
quadratic potential where they interact through Coulomb
interaction\cite{que92a}:
\begin{equation}
  \label{eq:SunNov13152833GMT2005}
H_{\mathrm{QD}}=\sum_{i=e,h}\left({\mathbf{p}_i^2\over2m_i}+{1\over2}m_i\omega^2\mathbf{r}_i^2\right)-{e^2\over\epsilon|\mathbf{r}_e-\mathbf{r}_h|}
\end{equation}
Here~$\mathbf{p}_i$ the momentum operator for the electron and
hole, $i=\mathrm{e}$, $\mathrm{h}$, respectively, $m_e$, $m_h$ the
electron and hole masses, $\omega$ the frequency which
characterises the strength of the confining potential, $e$ the
charge of the electron and $\epsilon$ the background dielectric
constant screening the Coulomb interaction. This hamiltonian
defines the two length scales of our problem, the 2D Bohr
radius~$a_\mathrm{B}$ and the dot size~$L$:
\begin{subequations}
  \label{eq:SunNov13154348GMT2005}
  \begin{eqnarray}
    a_\mathrm{B}&=&{\epsilon\hbar^2\over2\mu e^2}\\
    L&=&\sqrt{\hbar\over\mu\omega}
  \end{eqnarray}
\end{subequations}
where~$\mu=m_em_h/(m_e+m_h)$ is the reduced mass of the
electron-hole pair. To simplify the following discussion we assume
that~$m_e=m_h$, resulting in $\gamma_e=\gamma_h=\gamma$. The trial
wavefunction~(\ref{eq:MonNov14233031GMT2005}) separates
as~$\varphi(\mathbf r_e,
\mathbf{r}_h)=\mathcal{C}\Phi(\mathbf{R})\phi(\mathbf{r})$ where
$\mathbf{r}=\mathbf{r}_e-\mathbf{r}_h$ is the radius-vector of
relative motion and~$\mathbf{R}=(\mathbf{r}_e+\mathbf{r}_h)/2$ is
the center-of-mass position:
\begin{subequations}
  \label{eq:SunNov13194407GMT2005}
  \begin{eqnarray}
  \Phi(\mathbf{R})&=&{\sqrt{2(2\gamma+\gamma_{eh})}\over\sqrt{\pi}}
\exp{\left(-\mathbf{R}^2[2\gamma+\gamma_{eh}]\right)}
\label{eq:TueNov15000408GMT2005}\\
    \phi(\mathbf{r})&=&{\sqrt{2\gamma-\gamma_{eh}}\over\sqrt{2\pi}}
\exp{\left(-\mathbf{r}^2\left[{2\gamma-\gamma_{eh}\over4}\right]\right)}
\label{eq:TueNov15000413GMT2005}
  \end{eqnarray}
\end{subequations}

\begin{figure}
  \centering
  \includegraphics[width=\linewidth]{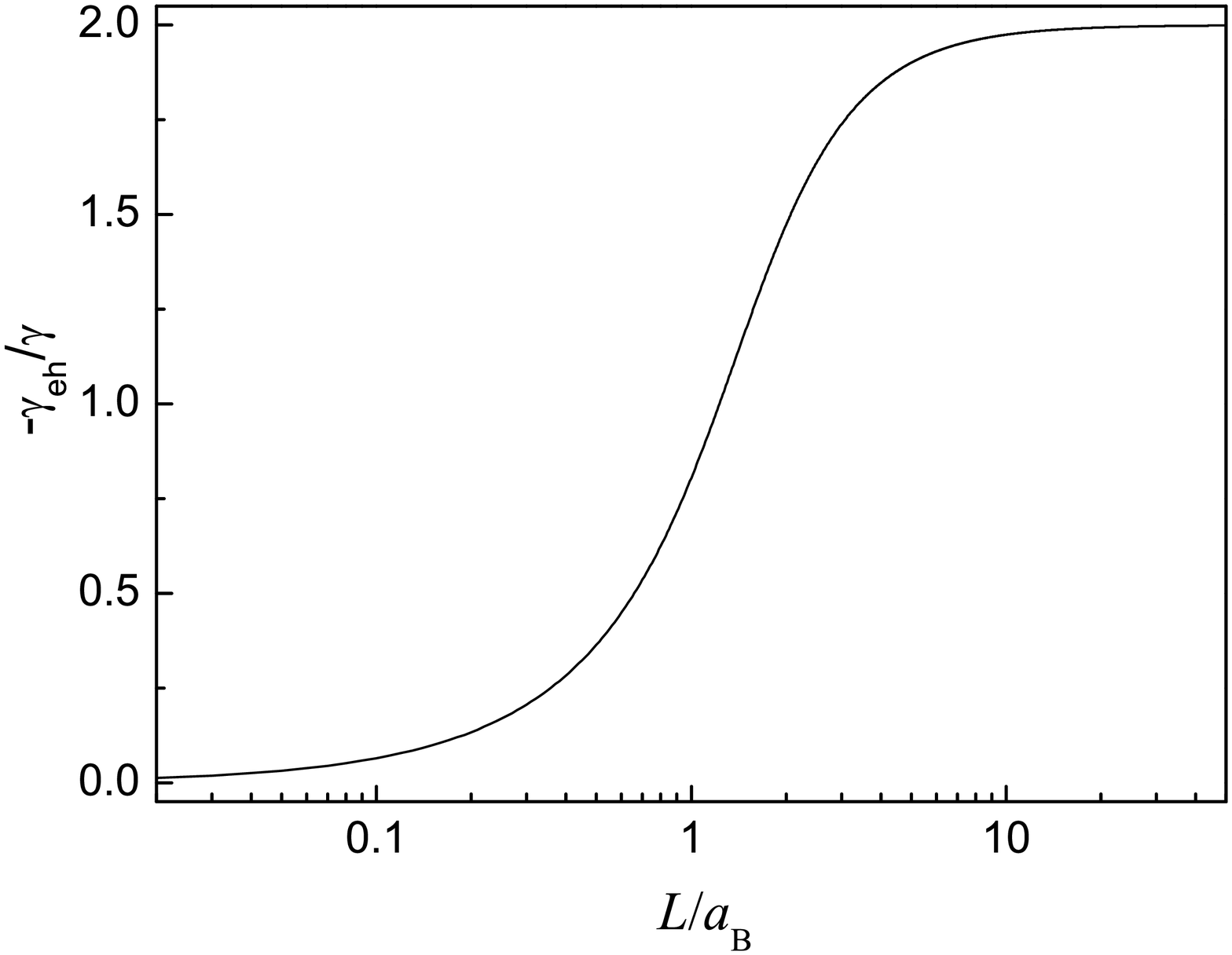}\\
  \caption{Ratio of parameters~$-\gamma_{eh}$ and $\gamma$
    (with~$\gamma=\gamma_e=\gamma_h$) as a function
    of~$L/a_\mathrm{B}$. For
    large dots where~$L\gg a_\mathrm{B}$,
    $-\gamma_{eh}/\gamma\approx2$ which corresponds to the bosonic
    limit where the electron and hole are strongly correlated.  For
    shallow dots where~$L\ll a_\mathrm{B}$,
    $-\gamma_{eh}/\gamma\approx0$ with electron and hole quantized
    separately. The transition is shown as the result of a variational
    procedure, with an abrupt transition when the dot size becomes
    comparable to the Bohr Radius.}
  \label{fig:TueNov15002111GMT2005}
\end{figure}

Eq.~(\ref{eq:TueNov15000408GMT2005}) is an eigenstate of the
center-of-mass energy operator and equating its parameters with those
of the exact solution yields the relationship
$2\gamma+\gamma_{eh}={2/L^2}$.  This constrain allows to
minimise~(\ref{eq:TueNov15000413GMT2005}) with respect to a single
parameter, $a=-\gamma_{eh}/2+1/(2L^2)$, which eventually amounts to
minimise ${4a_\mathrm{B}/a^2}+{a_\mathrm{B} a^2/L^4}-{2\sqrt\pi/a}$.
Doing so we have obtained the ratio~$-\gamma_{eh}/\gamma$ as a
function of~$L/a_\mathrm{B}$ displayed on
Fig.~\ref{fig:TueNov15002111GMT2005}. The transition from bosonic to
fermionic regime is seen to occur sharply when the dot size becomes
commensurable with the Bohr radius.
For large dots, i.e., for large values of~$L/a_\mathrm{B}$, the
ratio is well approximated by the expression
\begin{equation}
  \label{eq:TueNov15003541GMT2005}
  -\gamma_{eh}/\gamma = 2-(a_\mathrm{B}/L)^2
\end{equation}
so that in the limit~$a_\mathrm{B}/L\rightarrow0$,
Eq.~(\ref{eq:MonNov14233031GMT2005}) reads
$\varphi(\mathbf{r}_e,\mathbf{r}_h)\propto{\exp(-(\sqrt{\gamma_e}\mathbf{r}_e-\sqrt{\gamma_h}\mathbf{r}_h)^2)}$
with vanishing normalization constant. This mimics a free exciton
in an infinite quantum well. It corresponds to the bosonic case.
On the other hand, if $L$ is small compared to the Bohr radius,
with $\gamma_{eh}\rightarrow0$, the
limit~(\ref{eq:ThuAug11025623BST2005}) is recovered with
$\varphi\propto\exp(-\gamma_e\mathbf{r}_e^2)\exp(-\gamma_h\mathbf{r}_h^2)$.
This corresponds to the fermionic case.

One can readily check that~(\ref{eq:MonNov14233031GMT2005}) gives, in
the case $\gamma_{eh}\rightarrow-2\sqrt{\gamma_e\gamma_h}$, an exciton
binding energy which is smaller by only~$20\%$ than that calculated
with a hydrogenic wavefunction, which shows that the Gaussian
approximation should be tolerable for qualitative and
semi-quantitative results. Moreover, its form corresponds to the
general shape of a trial wavefunction for an exciton in an arbitrary
QD,~\cite{semina05a} with the only difference that we take the
Gaussian expression instead of a Bohr exponential for the wavefunction
of the relative motion of the electron and hole. This compromise to
numerical accuracy allows on the other hand to obtain analytical
expressions for all the key parameters, starting with the overlap
integrals~(\ref{eq:ThuAug11001646BST2005}) which take a simple form in
terms of multivariate Gaussians:
\begin{equation}
  \label{eq:TueJul12234613BST2005}
  \beta_m=\mathcal{C}^{2m}\int\exp(-{\bf x}^TA{\bf x})\,d{\bf x}\int\exp(-{\bf
y}^TA{\bf y})\,d{\bf y}
\end{equation}
where
\begin{subequations}
  \label{eq:ThuAug11142402BST2005}
  \begin{gather}
  \mathbf{x}^T=(x_{e_1},x_{e_2},\ldots,
x_{e_m},x_{h_1},x_{h_2},\ldots,x_{h_m})\label{eq:WedNov16162543GMT2005}\\
  \mathbf{y}^T=(y_{e_1},y_{e_2},\ldots,
y_{e_m},y_{h_1},y_{h_2},\ldots,y_{h_m})\label{eq:WedNov16162548GMT2005}
  \end{gather}
\end{subequations}
are the~$2m$ dimensional vectors which encapsulate all the
degrees of freedom of the~$m$ excitons-complex, and~$A$ is a
positive definite symmetric matrix which
equates~(\ref{eq:ThuAug11001646BST2005})
and~(\ref{eq:TueJul12234613BST2005}), i.e., which satisfies
\begin{multline}
  \label{eq:MonJul11160843BST2005}
  {\bf x}^TA{\bf x}=2\gamma_e\sum_{i=1}^m
x_i^2+2\gamma_h\sum_{i=m+1}^{2m}x_i^2+\gamma_{eh}x_mx_{m+1}\\+\gamma_{eh}\sum_{i=1}^mx_ix_{m+i}+\gamma_{eh}\sum_{i=1}^{m-1}x_ix_{m+i+1}
\end{multline}
and likewise for~$\mathbf{y}$ (to simplify notation we have not
written an index~$m$ on~$\mathbf{x}$, $\mathbf{y}$ and~$A$, but
these naturally scale with~$\beta_m$).  The identity for $2m$-fold
Gaussian integrals
\begin{equation}
  \label{eq:ThuAug11151254BST2005}
  \int\exp(-{\bf x}^TA{\bf x})\,d{\bf x}={\pi^{m}\over\sqrt{\det{A}}}
\end{equation}
allows us to obtain an analytical expression for $\beta_m$,
although a cumbersome one.  The determinant of the matrix $A$
reads
\begin{equation}
  \label{eq:MonJul18103640BST2005}
  \det{A}={\gamma_{eh}}^{2m}\sum_{k=0}^m\sum_{l=0}^{m-k}(-1)^{\lfloor
m/2\rfloor+k}\mathcal{A}_m(k,l)\left({\gamma_e\gamma_h\over{\gamma_{eh}}^2}\right)^k.
\end{equation}
Here we introduced a quantity
\begin{multline}
  \label{eq:SunJul17233433BST2005}
\mathcal{A}_m(k,l)=\mathcal{A}'_m(k,l)\\+\sum_{i=1}^m\left(\mathcal{A}'_{m-i}(k,l-i)-\mathcal{A}'_{m-i-1}(k,l-i)\right)
\end{multline}
and
\begin{multline}
  \label{eq:SunJul17175616BST2005}
\mathcal{A}'_m(k,l)=\sum_{\eta=1}^{p(l)}{(\sum_i\nu_\eta^l(i))!\over\prod_i\nu_\eta^l(i)!}\times\\{m-l\choose\sum_i\nu_\eta^l(i)}{m-l-\sum_i\nu_\eta^l(i)\choose
k-\sum_i\nu_\eta^l(i)}
\end{multline}
with~$k\in]0,m]$, $l\in[0,m]$ and~$p(l)$ and~$\nu_\eta(i)$ already
introduced as the partition function of~$l$ and the number of
occurence of~$i$ in its~$\eta$th partition. For the case~$k=0$ the
finite size of the matrix implies a special rule which
reads~$\mathcal{A}_m(0,l)=4\delta_{m,l}\delta_{m\equiv2,0}$.
Together with~(\ref{eq:ThuAug11141718BST2005}),
(\ref{eq:ThuAug11151254BST2005}) and
(\ref{eq:MonJul18103640BST2005}), expression
(\ref{eq:TueJul12234613BST2005}) provides the~$\beta_m$ in the
Gaussian approximation. One can see the considerable complexity of
the expressions despite the simplicity of the model wavefunction.
Once again, even a numerical treatment meets with difficulties
owing to manipulations of series of large quantities which sum to
small values. We had to turn to exact algebraic computations to
obtain $\alpha_n$ coefficients free from numerical artifacts.
Before we present the numerical results, we once again turn to the
limiting case of a large QD ($L \gg a_{\mathrm B}$) where the
bosonic behaviour may be expected, putting again for simplicity
$\gamma_e = \gamma_h = \gamma$.  When
Eq.~(\ref{eq:TueNov15003541GMT2005}) holds, rather lengthy
algebraic manipulations yield
\begin{equation}\label{betak_large}
  \beta_m \approx \frac{(2a_\mathrm{B}/L)^{2(m-1)}}{m^2}.
\end{equation}
This approximation is valid for small values of $a_\mathrm{B}/L$
and for $m\ll L/a_\mathrm{B}$. In computations of $\alpha_n$ with
$n\ll L/a_\mathrm{B}$, the denominator in Eq.~(\ref{betak_large})
plays a minor role and
\begin{equation}
  \label{alpha_n}
  \alpha_n = \sqrt{n}\sqrt{1- 2(n-1)(a_\mathrm{B}/L)^2}.
\end{equation}
Thus in the small-$n$ limit the matrix elements of the exciton
creation operator are close to that of bosons and the corrections
arise proportionally to the parameter $n(a_{\mathrm B}/L)^2$.

\begin{figure}[t]
\centering
\includegraphics[width=\linewidth]{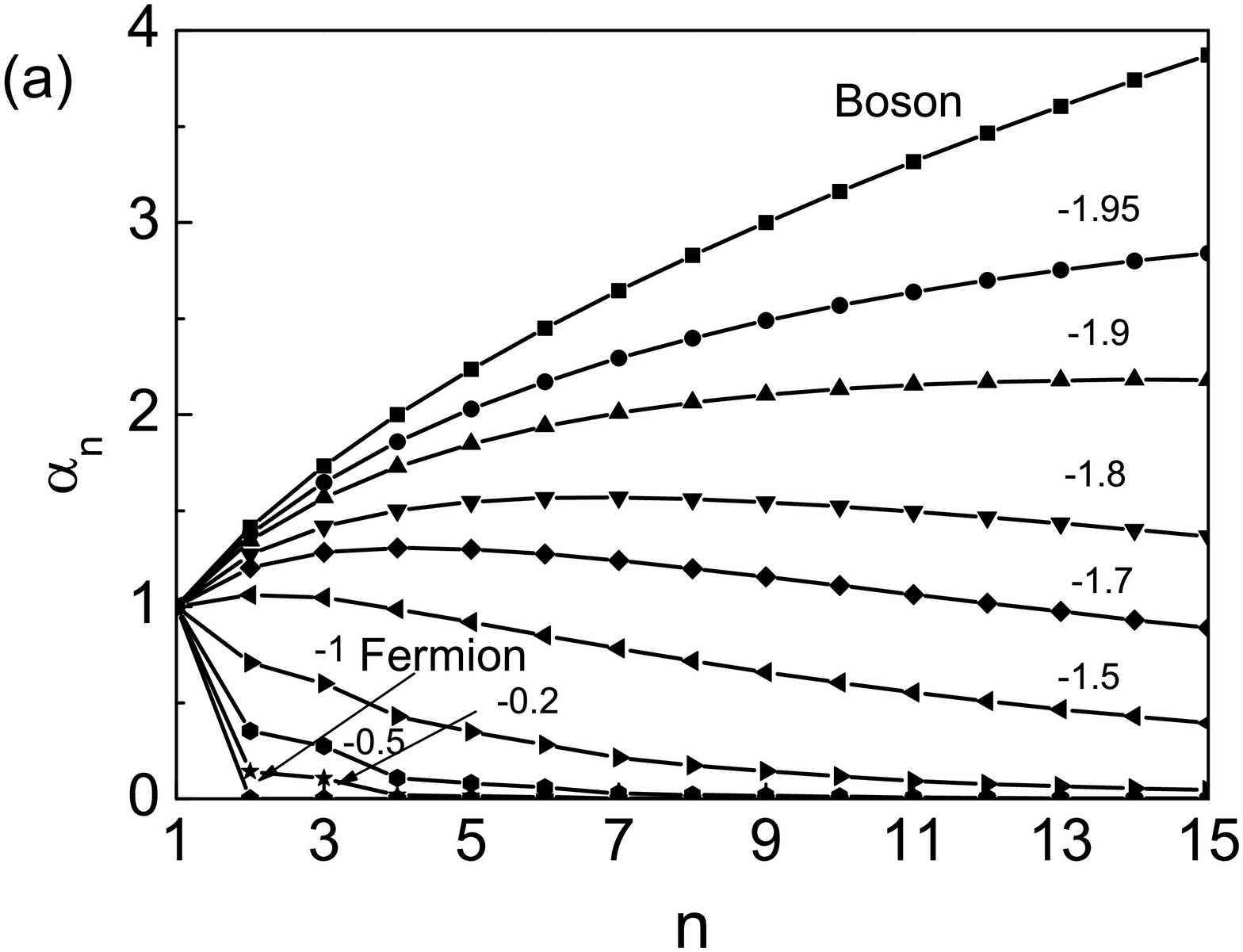}
\includegraphics[width=\linewidth]{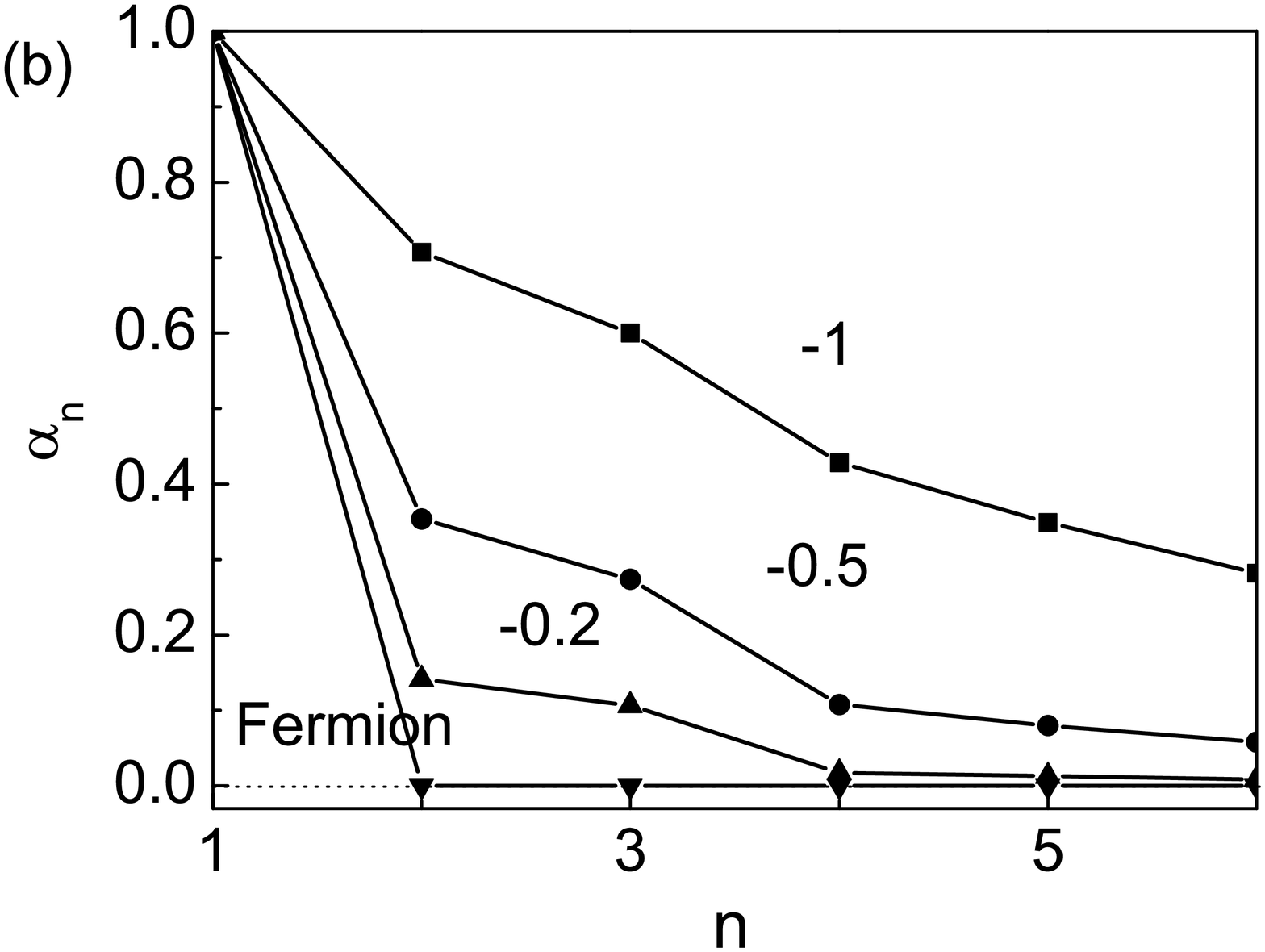}
\caption{(a) Matrix elements~$\alpha_n$ of the exciton creation
  operator $\ud{X}$ calculated for $n\le15$ for various trial
  wavefunctions. The top curve shows the limit of true bosons
  where~$\alpha_n=\sqrt n$ and the bottom curve the limit of true
  fermions where~$\alpha_n=\delta_{n,1}$.  Intermediate cases are
  obtained for values of~$\gamma_{eh}$
  from~$-1.95\sqrt{\gamma_e\gamma_h}$ down to
  $-0.2\sqrt{\gamma_e\gamma_h}$, interpolating between the boson and
  fermion limit. (b) Magnified region close to the fermion limit.
  Values displayed are everywhere given in units of
  $\sqrt{\gamma_e\gamma_h}$.}\label{fig:TueAug16170841BST2005}
\end{figure}

Fig.~\ref{fig:TueAug16170841BST2005} shows the behaviour of $\alpha_n$
for different values of $\gamma_{eh}$ interpolating from the bosonic
case ($\gamma_{eh} = -2\sqrt{\gamma_e\gamma_h}$) to the fermionic case
($\gamma_{eh} = 0$).  The crossover from bosonic to fermionic limit
can be clearly seen: for $\gamma_{eh}$ close to
$-2\sqrt{\gamma_e\gamma_h}$, the curve behaves like $\sqrt{n}$, the
deviations from this exact bosonic result becoming more pronounced
with increasing~$n$.  For $\gamma_{eh}$ close to
$-2\sqrt{\gamma_e\gamma_h}$, the curve initially behaves like
$\sqrt{n}$, the deviations from this exact bosonic result becoming
more pronounced with increasing~$n$.  The curve is ultimately
decreasing beyond a number of excitations which is smaller the greater
the departure of~$\gamma_{eh}$ from~$-2\sqrt{\gamma_e\gamma_h}$. After
the initial rise, as the overlap between electron and hole
wavefunctions is small and bosonic behaviour is found, the decrease
follows as the density becomes so large that Pauli exclusion becomes
significant. Then excitons cannot be considered as structure-less
particles, and fermionic characteristics emerge. With $\gamma_{eh}$
going to $0$, this behaviour is replaced by a monatonically decreasing
$\alpha_n$, which means that it is ``harder and harder'' to add
excitons in the same state in the QD; the fermionic nature of excitons
becomes more and more important.

\begin{figure}[t]
\centering
\includegraphics[width=\linewidth]{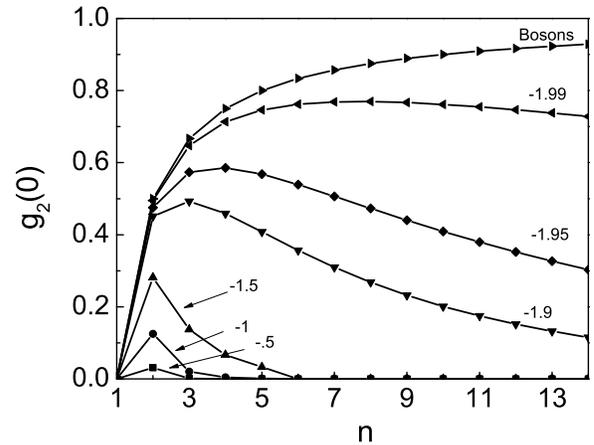}
\caption{Second order correlator $g_2(0)$ of the excitonic field
for
  Fock states~$\ket{n}$ given by
  Eqs.~(\ref{eq:WedAug10230402BST2005}--\ref{eq:WedAug10230706BST2005})
  as a function of~$\gamma_{eh}/\sqrt{\gamma_e\gamma_h}$ which
  interpolates between the fermionic ($-0.5$, $-1$ and~$-1.5$) and
  bosonic ($-1.9$, $-1.95$, $-1.99$) limits.  Upper line corresponds
  to pure bosons ($\gamma_{eh}=-2\sqrt{\gamma_e\gamma_h})$.  With
  increasing fermionic character the antibunching is very pronounced
  at high intensities. Close to the fermionic limit, $g_2$ is always
  very small and vanishes quickly.}\label{fig:WedSep14115006BST2005}
\end{figure}

An important quantity for single mode particles, especially in
connection to their coherent features, is the  (normalised) second
order correlator~$g_2$ which in our case reads
\begin{equation}
  \label{eq:WedSep14120534BST2005}
g_2(t,\tau)={\langle\ud{X}(t)\ud{X}(t+\tau)X(t+\tau)X(t)\rangle\over\langle\hat\mu(t)\rangle\langle\hat\mu(t+\tau)\rangle}
\end{equation}
where~$\hat\mu$ is the exciton number operator which satisfies
\begin{equation}
  \label{eq:SunAug14034111BST2005}
  \hat\mu\ket{n}=n\ket{n}
\end{equation}
with~$\ket{n}$ the bare exciton state with~$n$ electron-hole
pairs,
cf.~(\ref{eq:WedAug10230402BST2005}--\ref{eq:WedAug10230706BST2005}).
At our energy diagram level, we can only compute
zero-delay~$\tau=0$ correlations and there is no dynamics
so~$t\rightarrow\infty$. Eq.~(\ref{eq:WedSep14120534BST2005})
reduces to~$g_2(0)$ a quantity, which is one of great physical and
experimental relevance.  It will be sufficient for our description
to consider Fock states of excitons only, although an extension to
other quantum states is straightforward. The matrix representation
in the basis of states~$\ket{n}$ reads $X^{\dagger
  2}X^2=(\alpha_{i-1}^2\alpha_i^2\delta_{i,j})_{0\le i,j}$
with~$\alpha_i=0$ if~$i<1$, so that for a Fock state with~$n$
excitons,
\begin{equation}
  \label{eq:ThuNov10220650GMT2005}
  g_2(0)={\alpha_{n-1}^2\alpha_n^2\over n^2}
\end{equation}
Note that regardless of the model, $g_2(0)=0$ for~$n=1$. General
results are displayed on Fig.~\ref{fig:WedSep14115006BST2005}.
They correspond to the exciton field which is the one of most
interest to us, and could be probed in a two-photons correlation
experiment with the light emitted directly by the exciton. We
compare the result to the pure bosonic case
where~$\alpha_n=\sqrt{n}$ and therefore $g_2(0)=(n-1)/n$ so
that~$g_2\rightarrow 1$ with increasing number of particles which
expresses the similarity of an intense Fock state with a coherent
state (especially regarding their fluctuations). In our case,
however, the underlying fermionic structure results in an
\emph{antibunching} of excitons, i.e., the probability of finding
two excitons at the same time is lowered at high exciton
densities. Close to the fermionic limit, this antibunching is very
pronounced and it is very unlikely to have more than one exciton
in the system.

\section{Coupling to a single radiation mode of a
  microcavity}\label{sec:multiplets}

We now present the emission spectra of the coupled cavity-dot
system when the exciton field is described by the creation
operator~$\ud{X}$. The procedure is straight-forward in principle
and is a direct extension of the concepts discussed at length in
the introduction. The hamiltonian assumes the same form as
previously, but now with~$X$ as we defined it for the
matter-field~$b$:
\begin{equation}
  \label{eq:SunAug14021515BST2005}
  H=\hbar\omega(\ud{a}a+\hat\mu)+\hbar g(a\ud{X}+\ud{a}X)\,,
\end{equation}
As~(\ref{eq:SunAug14021515BST2005}) conserves the total number of
excitations~$\mu=n+m$, it can be decoupled by decomposition of the
identity as
\begin{equation}
  \label{eq:SunAug14034911BST2005}
\mathbf{1}=\sum_{\mu=0}^\infty\sum_{n=0}^{\min(N,\mu)}\ket{n,\mu-n}\bra{\mu-n,n}\,.
\end{equation}
where~$N$ is the smaller index~$n$ for which~$\alpha_n$ becomes
zero. In the Gaussian wavefunction approximation without
interaction, no~$\alpha_n$ ever becomes exactly zero, in which
case~$N\rightarrow\infty$ and the upper limit in the second sum of
Eq. (\ref{eq:SunAug14034911BST2005}) is $\mu$.

Inserting~(\ref{eq:SunAug14034911BST2005}) twice
in~(\ref{eq:SunAug14021515BST2005}) yields
$H=\bigoplus_{\mu=0}^\infty H_\mu$ with
\begin{multline}
  \label{eq:SunAug14034939BST2005}
  H_\mu=\hbar\omega\mu+{}\\\hbar
g[\alpha_n\sqrt{\mu-n+1}\delta_{n,m+1}+\mathrm{h.c.}]_{1\le
n,m\le\min(N,\mu)}
\end{multline}
in the basis of bare photon-exciton states:
\begin{multline}
  \label{eq:SunAug14035120BST2005}
\mathcal{H}_\mu=\{\ket{0,\mu},\ket{1,\mu-1},\ldots,\\\ket{\min(N,\mu),\mu-\min(N,\mu)}\}\,.
\end{multline}
If~$\alpha_n$ does not vanish, the basis has~$\mu+1$ states in
this manifold with last state~$\ket{\mu,0}$ having all photons
transferred in the excitonic field. In the case where~$\alpha_n$
vanishes, the excitonic field saturates and the further
excitations are constrained to remain in the photonic field.

Following the nomenclature laid down in the introduction, we write
the dressed state~$\ketd{\nu,\mu}$ for the~$\nu^\mathrm{th}$
eigenstate of the manifold with~$\mu=n+m$ excitations ($n$
excitons + $m$ photons) and
$c_n^{\nu,\mu}=\braketd{n,\nu-n}{\nu,\mu}$ its decomposition on
bare states of this manifold, i.e.,
\begin{equation}
  \ketd{\nu,\mu}=\sum_{n=0}^{\min(N,\mu)}c_n^{\nu,\mu}\ket{n,\mu-n}
\end{equation}
We compute the emission spectra corresponding to transitions
between multiplets, with matrix elements
$I_\mathrm{end}=|\brad{\nu',\mu-1}a\ketd{\nu,\mu}|^2$ for emission
of a photon from the cavity
and~$I_\mathrm{lat}=|\brad{\nu',\mu-1}X\ketd{\nu,\mu}|^2$ for
direct exciton emission into a non-cavity mode. Then
\begin{gather}
  \label{eq:SunAug14040622BST2005}
    I_\mathrm{end}=\left|\sum_{n=0}^{\min(N,\mu-1)}(c_n^{\nu',\mu-1})^*
c_n^{\nu,\mu}\sqrt{\mu-n}\right|^2\,,\\
    I_\mathrm{lat}=\left|\sum_{n=1}^{\min(N,\mu)}(c_{n-1}^{\nu',\mu-1})^*
c_n^{\nu,\mu}\alpha_n\right|^2\,.
\end{gather}
In cavity QED terminology, $I_\mathrm{end}$ and~$I_\mathrm{lat}$
correspond to \emph{end--emission} and~\emph{lateral--emission}
photo--detection respectively, while in luminescence of a
microcavity, one observes the linear combination of the two
contributions simultaneously. $I_\mathrm{lat}$ reflects most the
behaviour of the excitonic field. To separate the two, one can
make use of the scattering geometry; in a pillar structure, for
instance, the cavity photon emission is predominantly through the
end mirrors, so $I_\mathrm{lat}$ could be detected on the edge of
the structure. These measurements display the most interesting
features and we focus on them.

\begin{figure}[t]
      \includegraphics[width=0.8\linewidth]{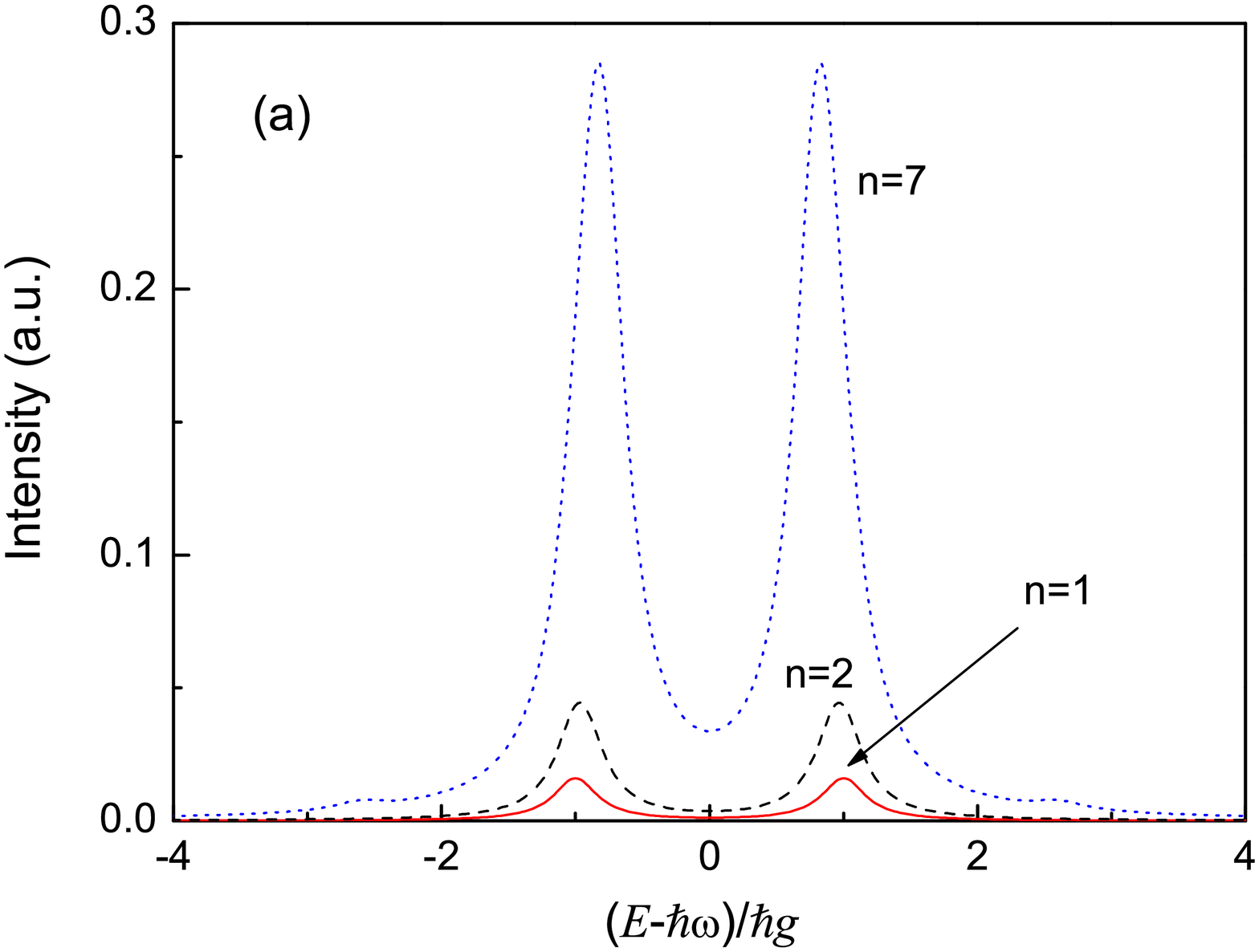}
      \includegraphics[width=0.8\linewidth]{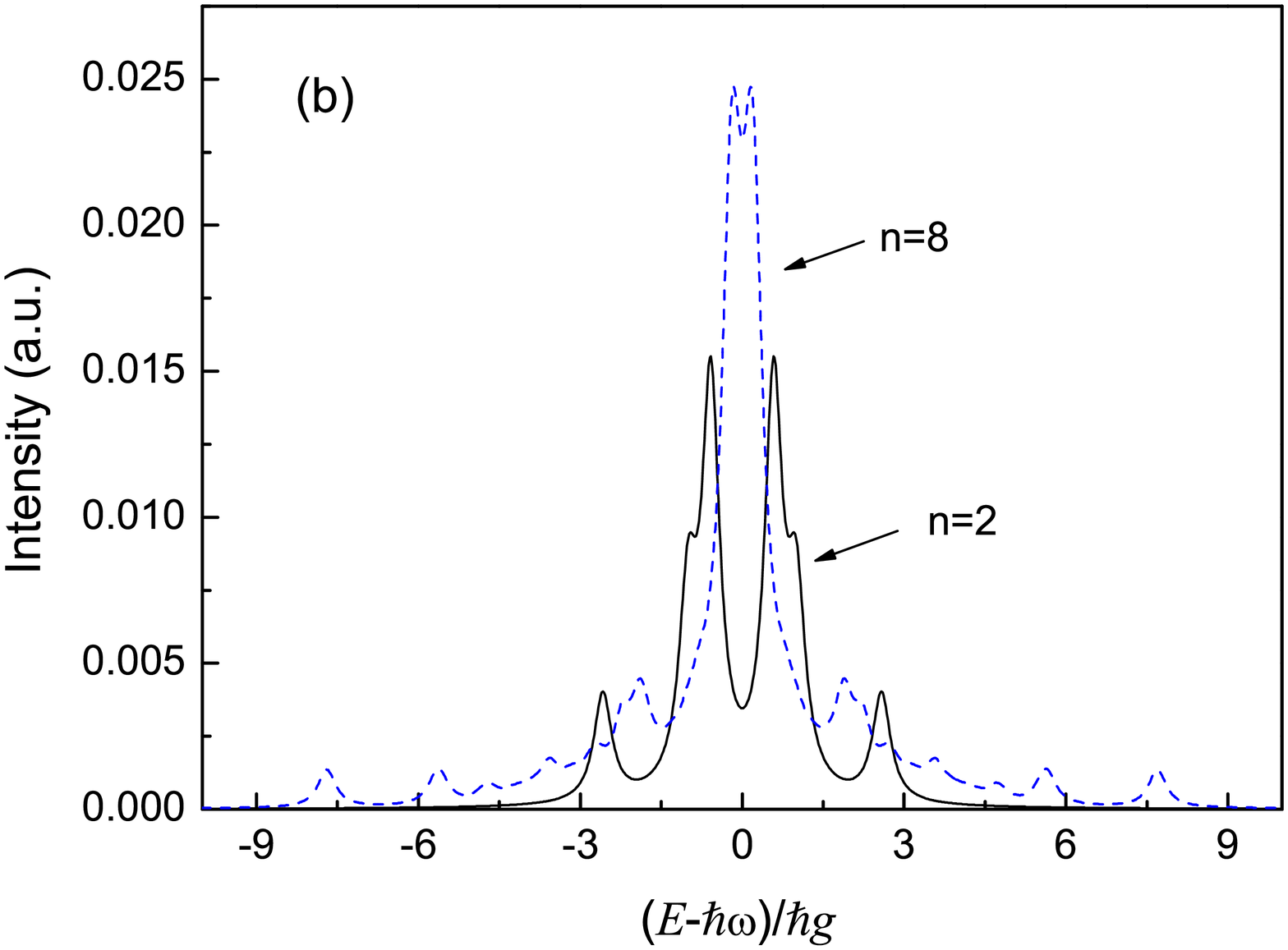}
      \includegraphics[width=0.8\linewidth]{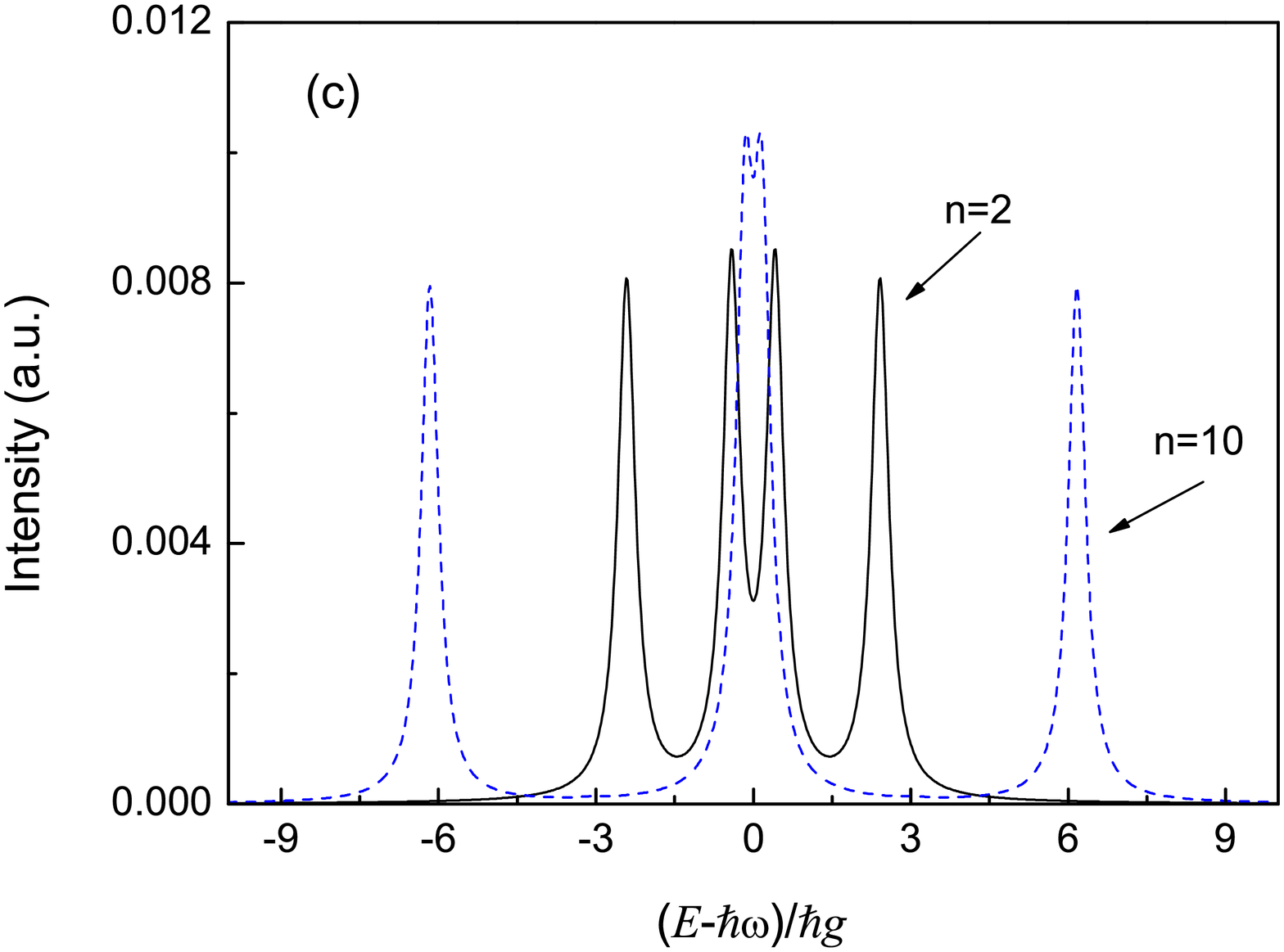}
      \caption{(Color online) Spectra for various intensities of the light field at
        specific values of~$\gamma_{eh}$: (a) close to the bosonic
        limit with~$\gamma_{eh}=-1.9\sqrt{\gamma_e\gamma_h}$
        for~$n=1$ (solid red), $2$ (dashed blue) and $7$ (dotted black)
        featuring a broadened and red-shifted
        Rabi doublet as the intensity increases and the onset of a
        multiplet structure, (b) intermediate case:
        $\gamma_{eh}=-\sqrt{\gamma_e\gamma_h}$ for $n=2$ (solid black) and
        $10$ (dashed blue)
        demonstrating a complicated multiplet structure and (c) close
        to the fermionic case with
        $\gamma_{eh}=-0.05\sqrt{\gamma_e\gamma_h}$ for $n=2$ (solid black) and
        $10$ (dashed blue)
        featuring quadruplet structure going towards Mollow triplet at
        high intensities.}\label{fig:multiplets}
\end{figure}

Fig.~\ref{fig:multiplets} shows the calculated emission spectra
$I_\mathrm{lat}$ of a QD embedded in a cavity. All the spectra are
broadened by convolution with a Lorentzian of width $\gamma =0.2
g$. Figs.~(a), (b) and (c) are, respectively, the results close to
the bosonic limit ($\gamma_{eh}=-1.99\sqrt{\gamma_e\gamma_h}$), in
between ($\gamma_{eh}=-\sqrt{\gamma_e\gamma_h}$) and close to the
fermionic limit ($\gamma_{eh}=-0.05\sqrt{\gamma_e\gamma_h}$). Each
curve is labelled with the number of excitations in the manifold,
reflecting the intensity of the pumping field.
Fig.~\ref{fig:multiplets}(a) shows a pronounced Rabi doublet in
the case $n=1$ (vacuum field Rabi splitting) in accordance with
the general theory set out in the introduction.  Higher manifolds
reveal non-bosonic behaviour, with a reduced Rabi splitting
doublet when $n=2$, and the onset of a multiplet structure when
$n=5$, where small peaks appear at $E \approx \hbar(\omega\pm 2
g)$. The increase of dot confinement in the case
$\gamma_{eh}=-\sqrt{\gamma_e\gamma_h}$ in
Fig.~\ref{fig:multiplets}(b), makes deviations from bosonic
behaviour more pronounced, so even for $n=2$ a multiplet structure
is observable.  For higher manifold numbers, more lines appear,
and a decrease of the splitting of the central Rabi doublet
occurs.  Further strengthening the confinement to $\gamma_{eh} =
-0.05\sqrt{\gamma_e\gamma_h}$ makes the manifestations of
fermionic behaviour very clear; for $n=2$, a quadruplet structure
is seen, with the central two peaks almost merged.  The situation
is similar to the Mollow triplet of the exact fermionic limit,
even more so for an increase of manifold number from $n=2$ to
$n=10$, where the separation of the side peaks grows and the
central ones merge further. It can be seen from
Fig.~\ref{fig:multiplets} that the decrease of the QD size and the
corresponding change of the exciton quantisation regime
(with~$\gamma_{eh}$ ranging from $-2\sqrt{\gamma_e\gamma_h}$ to
$0$) is manifest in the emission spectra as a transition from a
Rabi doublet to a Mollow triplet. The increase of the excitation
power (at fixed QD size) leads qualitatively to the same effect;
fermionic behaviour becomes more pronounced with the increase of
the excitation power and decrease of the QD size.

\begin{figure}[t]
\centering
\includegraphics[width=\linewidth]{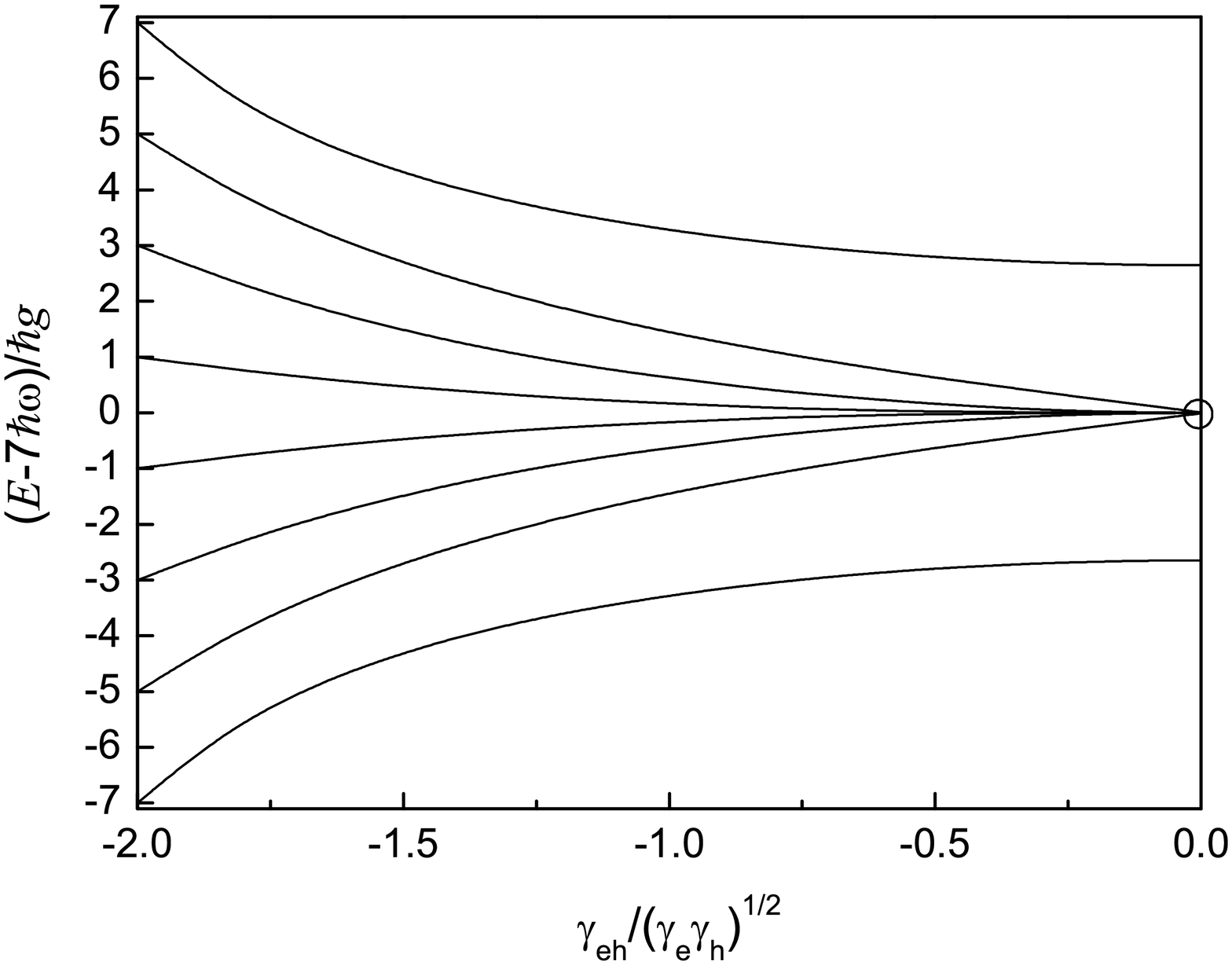}
\caption{Energy diagram of the dressed excitons over the
  interval~$\gamma_{eh}\in[-2\sqrt{\gamma_e\gamma_h},0]$ for the 7th
  manifold. The extremal values recover the diagrams of
  Fig.~\ref{fig:TueAug16164523BST2005} with~$n+1$ equally spaced
  energy levels in $n$th dressed boson manifold (far left) and twofold
  energy diagram with square root splitting for dressed fermions (far
  right).  The two outer boson energies connect smoothly to the two
  fermion energies while the~$n-1$ other one degenerate into a central
  line which disappears right when the system hits the Fermi limit
  (this point, $\gamma_{eh}=0$, is shown by
  circle).}\label{fig:WedSep14114927BST2005}
\end{figure}

Having demonstrated how our model interpolates between Fermi and
Bose statistics, we now investigate in further detail the
intermediate regime, by following the evolution of a single
manifold between the two limits. In
Fig.~\ref{fig:WedSep14114927BST2005} the eigenvalues are displayed
for the~$7$th manifold, also shown on
Fig.~\ref{fig:TueAug16164523BST2005} for bosons and fermions.
Eigenvalues are plotted as a function of~$\gamma_{eh}$ running
from~$-2\sqrt{\gamma_e\gamma_h}$, which recovers the left hand
side of Fig.~\ref{fig:TueAug16164523BST2005}, to~$0$, which
recovers the right hand side.  Other manifolds behave in
qualitatively the same way. In all cases, the~$n+1$ equally spaced
energies of the dressed bosons link to the two energies of the
dressed fermions as follows: the upper
energy~$\mu\hbar\omega+\mu\hbar g$ of the Bose limit links to the
upper energy~$\mu\hbar\omega+\sqrt{\mu}\hbar g$ of the Fermi
limit, and the symmetric behaviour occurs with the lower limit,
linking~$\mu\hbar\omega-\mu\hbar g$
with~$\mu\hbar\omega-\sqrt{\mu}\hbar g$. More interestingly, the
intermediate energies degenerate from the equal spacing of the
Bose limit towards~$\mu\hbar\omega$ near the Fermi limit, with a
discontinuity when~$\alpha_2$ becomes zero and only two
eigen-energies remain. Physically, this discontinuity arises
because for any $\gamma_{eh} \ne 0$, formally, an infinite number
of excitations can be fitted in the QD, i.e. $\alpha_n$ never
becomes exactly zero, see Fig.~\ref{fig:TueAug16170841BST2005}.
With $\gamma_{eh}$ approaching zero these modes couple to light
more and more weakly, thus the energies of the dressed states
become close to the bare energy $\mu \hbar \omega$. When
$\gamma_{eh}=0$ the fermionic regime is recovered and only two
dressed eigenstates with energies $\mu \hbar\omega \pm
\sqrt{\mu}\hbar g$ survive.

Although the eigenvalue structure of the dressed excitons in the
general case is simple and uniform, as is seen in
Fig.~\ref{fig:WedSep14114927BST2005}, the multiplet structures
which arises from it is, as has been seen in the spectra
previously investigated, rich and varied.  In the general case
where $-2\sqrt{\gamma_e\gamma_h} < \gamma_{eh} < 0$, two adjacent
manifolds contain respectively $n+1$ and $n$ levels, transitions
between any pair of which are possible. One may expect $n(n+1)$
lines in the emission spectra with a range of intensities and
positions. The phenomenological broadening we have introduced
leads to the decrease of the number of resolvable lines.  In all
cases the $n=1$ to $n=0$ transition provides only two lines, with
constant energies throughout, because with only one exciton
present the question of bosonic or fermionic behaviour is
irrelevant. To access the consequences of exciton statistics, one
must, unsurprisingly, reach higher manifolds.

\begin{figure}[t]
  \centering \includegraphics[width=\linewidth]{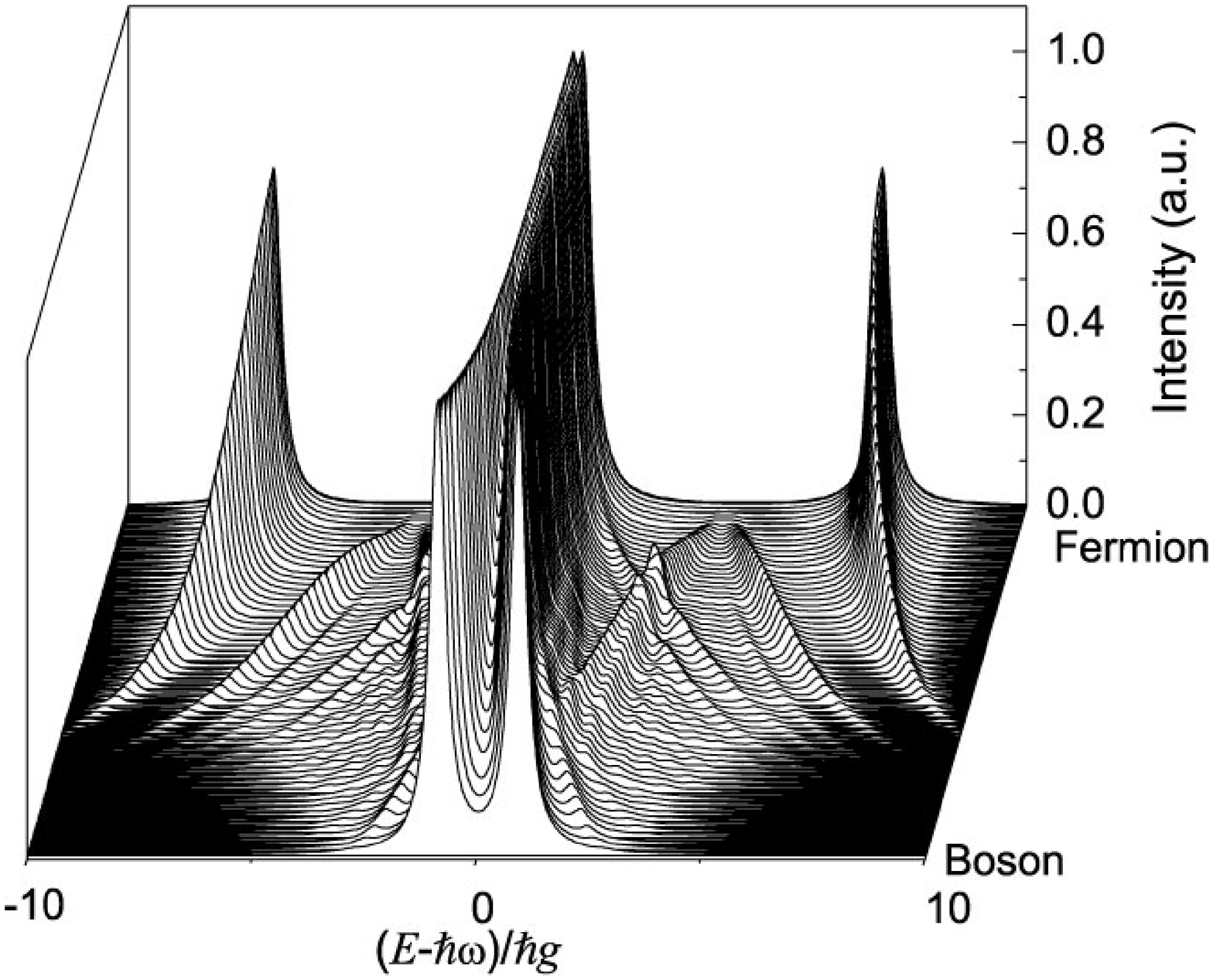}
  \includegraphics[width=\linewidth]{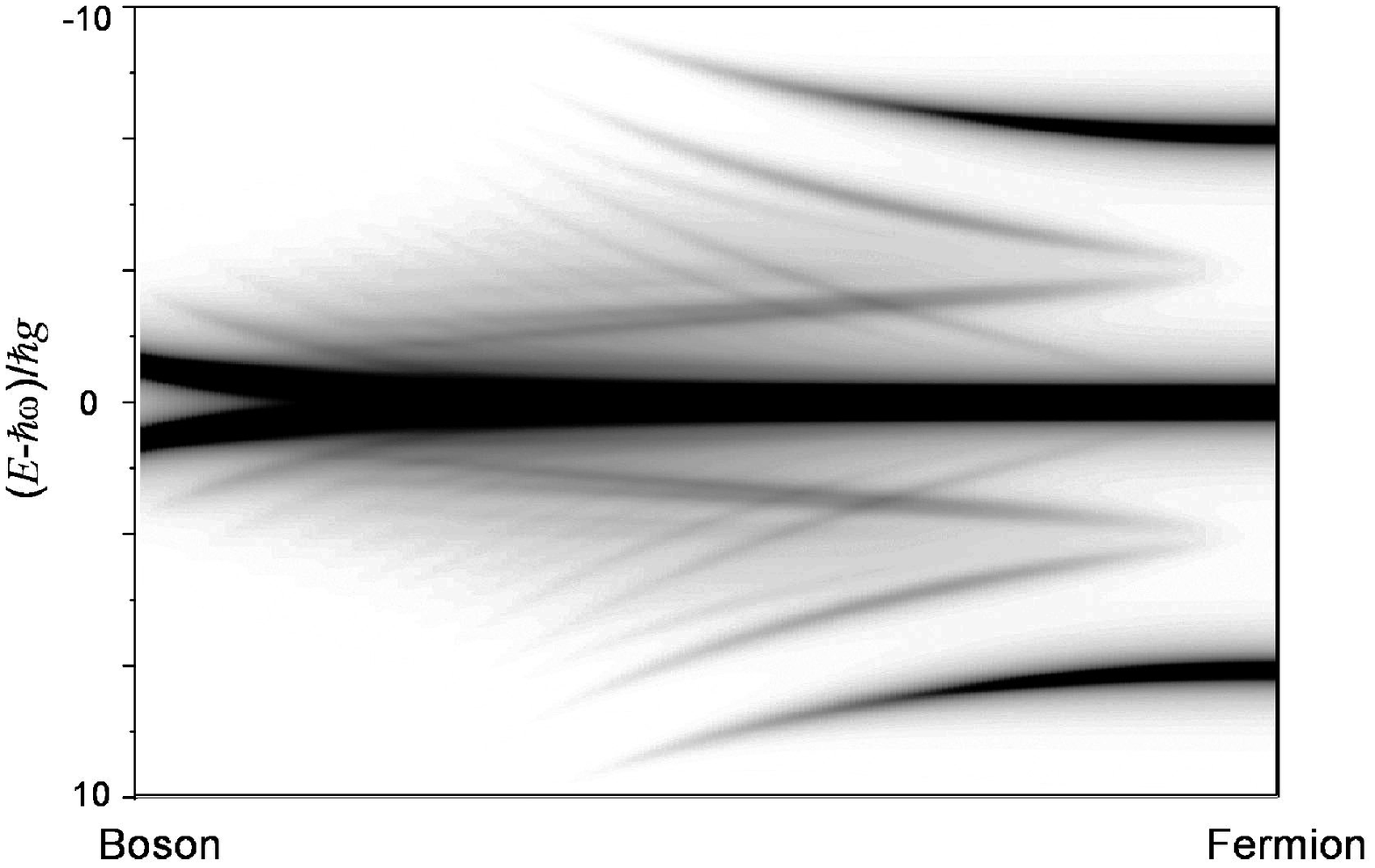}
  \caption{Transition from Fermi to Bose limits as observed in the
  optical emission spectra. (a) Superpositions of spectra
  for~$\gamma_{eh}=-2\sqrt{\gamma_e\gamma_h}$ in front of the figure
  to~$\gamma_{eh}=0$ at the back, recovering respectively the Rabi
  doublet and Mollow triplet. In the intermediate region, intricate
  and rich patterns of peaks appear, split, merge, or disappear.  (b)
  Same data as in (a), but as a density plot on logarithmic
  scale, to discriminate the peaks, their positions and splitting as
  well as their relative intensities. All spectra are normalized to
  maximum intensity equal to $1$.}  \label{fig:WedAug17025302BST2005}
\end{figure}

In Fig.~\ref{fig:WedAug17025302BST2005} we show the evolution of
the spectra between the boson and fermion limits. They two panels
are different projections of the same data, namely the multiplet
structure as a function
of~$\gamma_{eh}\in[-2\sqrt{\gamma_e\gamma_h},0]$ for $n=12$. The
Rabi doublet is seen to evolve into a Mollow triplet going through
a complex and intertwined set of peaks whose splitting and
relative heights vary with the value of~$\gamma_{eh}$ considered.
The spectrum obtained is therefore a direct probe of the
underlying exciton quantum statistics.

As a final comment, we note that if Eq.~(\ref{alpha_n}) was to
hold for all~$n$, it would yield, apart from a renormalization by
$\sqrt{N}$, the  Dicke model~\cite{dicke54a}, which has been
widely used to describe various strong light-matter coupling
phenomena~\cite{eastham01a, keeling04a}. In our model it is
recovered when~$\beta_m = 1/N^{(m-1)}$, a case that we have
investigated in Ref.~[\onlinecite{laussy05d}].

At the heart of Dicke model lies the creation operator~$J_+$ for
an excitation of the ``matter field'' which distributes the
excitation throughout the assembly of~$N$ identical two-levels
systems described by fermion operators~$\sigma_i$, so that
$b^\dag$ in Eq.~(\ref{eq:TueAug9172340BST2005}) maps to~$J_+$ with
\begin{equation}
  \label{eq:TueAug9164730BST2005}
  J_+=\sum_{i=1}^N\sigma_i^\dag
\end{equation}
One checks readily that~$J_+$ and~$J_-=\ud{J_+}$ thus defined obey
an angular momentum algebra with magnitude~$N(N+1)$ (and
maximum~$z$ projection of~$J_z$ equal to $N$). In this case the
Rabi doublet arises in the limit where the total number of
excitations~$\mu$ (shared between the light and the matter field)
is much less than the number of atoms, $\mu\ll N$, in which case
the usual commutation relation~$[J_-,J_+]=-2J_z$ becomes
$[J_-/\sqrt{N},J_+/\sqrt{N}]\approx1$, which is the commutation
for a bosonic field. This comes from the expression of a Dicke
state with~$\mu$ excitations shared by~$N$ atoms given as the
angular momentum state~$\ket{-N/2+\mu}$. Therefore the
annihilation/creation operators~$J_-$, $J_+$ for one excitation
shared by~$N$ atoms appear in this limit like renormalised bose
operators~$\sqrt N a$, $\sqrt N\ud{a}$, resulting in a Rabi
doublet of splitting~$2\hbar g\sqrt N$. Such a situation
corresponds, e.g., to an array of small QDs inside a microcavity
such that in each dot electron and hole are quantized separately,
while our model describes a single QD which can accommodate
several excitons.  The corresponding emission spectra are close to
those obtained here below the saturation limit $\mu \ll N$, while
the nonlinear regime $N \gg 1$, $\mu \gg 1$ has peculiar
behaviour, featuring non-lorentzian emission line shapes and a
non-trivial multiplet structure, like the ``Dicke
fork''.~\cite{laussy05d}

\section{Conclusions and Prospects}

The spectrum of light emitted by quantum dot excitons in leaky
modes of a microcavity---which could be typically lateral
emission---is a signature of the quantum statistics of excitons. A
multiplet structure has been theoretically predicted with various
features which can help identify the exciton field statistics. We
provided the formalism to obtain the spectra expected for a
general dot in various geometries based on the form of the single
exciton wavefunction.  We investigated a generic case analytically
through a Gaussian approximation. The richness and specificity of
the resulting spectra provides a means to determine, through the
peak splittings and strength ratios, the parameter $\gamma_{eh}$
that measures how close excitons are to ideal fermions or bosons.
Although heavy numerical computations are required for realistic
structures, physical sense motivates that in small dots a
Fermi-like behaviour of excitations with separately quantised
electrons and holes is expected, while in larger dots behaviour
should converge towards the Bose limit. These trends should be
observable in the nonlinear regime (where more than one exciton
interacts at a single time with the radiation mode): depending on
whether Rabi splitting is found, or if a Mollow triplet or a more
complicated multiplet structure arises, one will be able to
characterize the underlying structure of the exciton field.

We have left the discussion of the interaction between excitons
out of the scope of the present paper considering the effects
which arise solely from Pauli exclusion principle. The effect of
interactions, through screening of Coulomb potential by the
electron-hole pairs, has been discussed in
Refs.~[\onlinecite{rombouts05a, tanguy02a}].  These papers
demonstrate that the interactions will make fermionic behaviour
more pronounced, since presence of other excitons screens the
Coulomb interaction which binds electron and hole and leads to the
increase of the Bohr radius. The question whether interactions
will predominate over Pauli exclusion lies beyond the scope of
this paper and we postpone it for a future work. Our preliminary
estimations show that even at moderate excitonic density the
effect of screening is small and does not lead to strong
qualitative deviations.\footnote{The main departure of our
approach from that of Refs.~[\onlinecite{rombouts05a, tanguy02a}]
is that we consider the screening by neutral excitons while they
take into account the screening by an electron-hole plasma. The
former approach results in a much weaker screening.}

To summarise, we have studied the effect of Pauli exclusion on the
optical emission spectra of microcavities with embedded QDs in the
strong coupling regime. We derived general expressions for the
exciton creation operator which allow systematic computation of
the light-matter coupling.  The crossover between bosonic
behaviour---observed in large QDs as Rabi doublet---to the
fermionic behaviour---observed in small QDs as Mollow
triplet---has been demonstrated.

\begin{acknowledgments}
  The authors thank Dr.~Ivan Shelykh, Dr.~Yuri Rubo and Marina~A.~Semina
  for helpful discussions.  M.M.G.  acknowledges the financial support
  by RFBR and ``Dynasty'' foundation---ICFPM. This work was supported
  by the Clermont-2 project MRTN-CT-2003-S03677.
\end{acknowledgments}

\bibliography{physrev}

\end{document}